\documentclass[a4paper,11pt]{article}
\pdfoutput=1 
\usepackage{amsfonts}
\usepackage{amsmath}
\usepackage{jheppub}
\usepackage[utf8]{inputenc}
\usepackage{booktabs}
\usepackage{multirow}
\usepackage{graphicx}
\usepackage{xcolor}
\definecolor{ochre}{RGB}{176,128,32}
\definecolor{revisionbrown}{RGB}{139,69,19}
\definecolor{beanpastegreen}{RGB}{75,130,90}

\makeatletter

\begin{document}

\title{Learning Effective Bulk Fields from Fermionic Spectra: Application to an Extended Power-Law Liquid Model for Cuprate Strange Metals}

\author[a]{Hong-Zhi Xiao,}
\author[a,b]{Zhan-Zhi He,}
\author[b,c]{Zhuo-Yu Xian,}
\author[a,d]{Shao-Feng Wu}

\affiliation[a]{Department of Physics, Shanghai University, Shanghai, 200444, China}
\affiliation[b]{School of Physics and Astronomy, Sun Yat-Sen University, 2 Daxue Road, Zhuhai 519082, China}
\affiliation[c]{Department of Physics, Freie Universit\"at Berlin, Arnimallee 14, DE-14195 Berlin, Germany}
\affiliation[d]{Center for Gravitation and Cosmology, Yangzhou University, Yangzhou 225009, China}

\emailAdd{hongzhixiao@shu.edu.cn}
\emailAdd{hezhzh29@mail2.sysu.edu.cn}
\emailAdd{xianzhy5@mail.sysu.edu.cn}
\emailAdd{sfwu@shu.edu.cn}

\abstract{
We develop a Neural-ODE framework to infer effective bulk metric functions and the charge-weighted gauge potential from boundary fermionic spectra within a static, planar black-hole ansatz. Controlled benchmarks recover the known bulk profiles of the Einstein--Maxwell and Gubser--Rocha models. We then use normalized interaction self-energy targets generated from an extended power-law liquid model with parameters calibrated by angle-resolved photoemission measurements to test the proposed geometric explanation of cuprate strange-metal spectra in a semi-holographic setting. For this application, we use a massless probe in backgrounds with conformal-to-$\mathrm{AdS}_2\times\mathbb{R}^2$ asymptotics and derive a spectral extraction formula from the conserved radial Dirac flux. We identify coordinate/Weyl redundancy and a degeneracy in the scaled Hawking temperature of the normalized inverse problem. The conformal factor remains undetermined, so bulk entropy and the associated contribution to electronic specific heat require independent input. Low-temperature targets are fitted accurately across the studied doping range. In the high-temperature test, however, the present ansatz and optimization search have not yielded a refined geometry that accurately reproduces the thermal spectral shape while preserving frequency symmetry. These results may guide the further development of semi-holographic models for cuprates, and the framework may also find applications in other strongly correlated electron systems.
}

\maketitle
\section{Introduction}\label{sec:intro}

The strange-metal regime of cuprate high-temperature superconductors exhibits striking departures from conventional Fermi-liquid behavior~\cite{landauTheoryFermiLiquid1959,Keimer:2015zqv,phillipsStrangerMetals2022}. Key signatures include linear-in-temperature resistivity $\rho \propto T$~\cite{gurvitchResistivity1825Sr1987}, a Hall-angle cotangent $\cot\theta_H \propto T^2$ suggestive of distinct longitudinal and Hall relaxation rates~\cite{chienEffectZnImpurities1991,carringtonTemperatureDependenceHall1992}, and a logarithmically enhanced specific-heat coefficient $C/T \propto \ln(1/T)$ near quantum criticality~\cite{michonThermodynamicSignaturesQuantum2019,girodNormalStateSpecific2021,lohneysenFermiliquidInstabilitiesMagnetic2007}. Of these signatures, the $T$-linear resistivity is often associated with the Planckian scattering rate, $\hbar/\tau\sim k_B T$~\cite{Hartnoll:2021ydi,zaanenWhyTemperatureHigh2004,patelTheoryPlanckianMetal2019}. Correlations between its coefficient and $T_c$ across several unconventional superconductors further suggest an intimate link between strange metallicity and superconductivity~\cite{jinLinkSpinFluctuations2011,yuanScalingStrangemetalScattering2022}. To understand these mysteries, various theoretical frameworks have been proposed, including quantum critical scenarios~\cite{phillipsBreakdownOneParameterScaling2005,huangDisorderedQuantumCritical2023}, the marginal Fermi liquid theory~\cite{varmaPhenomenologyNormalState1989,vallaEvidenceQuantumCritical1999}, and solvable SYK-type models~\cite{patelUniversalTheoryStrange2023,guoLarge$N$TheoryCritical2022,liStrangeMetalSuperconductor2024,chowdhurySachdevYeKitaevModelsWindow2022}, yet a complete microscopic mechanism remains elusive.

AdS/CFT correspondence~\cite{maldacenaLargeLimitSuperconformal1999,gubserGaugeTheoryCorrelators1998,wittenSitterSpaceHolography1998} offers a non-perturbative approach to strongly correlated systems by mapping a $d$-dimensional conformal field theory to a $(d+1)$-dimensional gravitational theory in anti-de Sitter (AdS) spacetime, where the radial direction encodes the renormalization group flow~\cite{Freedman:1999gp,deBoer:1999tgo,Skenderis:2002wp,Heemskerk:2010hk}. The strong-weak nature of the duality has enabled holographic methods to address problems in condensed matter~\cite{zaanenHolographicDualityCondensed2015,hartnollHolographicQuantumMatter2018}, QCD~\cite{babingtonChiralSymmetryBreaking2004,kruczenskiHolographicDualLargeN_c2004,Kim:2012ey}, and quantum information~\cite{rangamaniHolographicEntanglementEntropy2017,Chen:2021lnq}. Combined with linear response theory~\cite{kuboStatisticalMechanicalTheoryIrreversible1957}, holographic methods have yielded key results for fermionic response in charged black hole backgrounds~\cite{Lee:2008xf,Cubrovic:2009ye,liuNonFermiLiquidsHolography2011,iqbalRealtimeResponseAdS2009,faulknerEmergentQuantumCriticality2011,Faulkner:2010da,faulknerStrangeMetalTransport2010,faulknerHolographicNonFermiLiquid2011,faulknerChargeTransportHolographic2013}. The Reissner-Nordstr\"om-AdS (RN-AdS) black hole is a natural arena, as its low-temperature near-horizon geometry develops an $\mathrm{AdS}_2 \times \mathbb{R}^2$ factor that furnishes a gravitational realization of quantum criticality~\cite{faulknerEmergentQuantumCriticality2011,Iqbal:2011aj}. Among the various holographic constructions, the Gubser--Rocha model~\cite{gubserPeculiarPropertiesCharged2010}---an Einstein--Maxwell-dilaton (EMD) theory---has attracted a lot of interest: it allows for an analytical black hole solution, its dilaton coupling produces a vanishing ground-state entropy, and its near-horizon geometry remains conformal to $\mathrm{AdS}_2 \times \mathbb{R}^2$~\cite{gouterauxChargeTransportHolography2014,balmLinearResistivityOptical2023}, successfully reproducing $\rho \propto T$~\cite{davisonHolographicDualityResistivity2014,jeongLinear$T$ResistivityHigh2018,ahnLinear$T$ResistivityLow2020}. Blake and Donos~\cite{Blake:2014yla} further argued within EMD-Axion theory that both $\rho\propto T$ and $\cot \theta_H\propto T^2$ can be obtained simultaneously, a conclusion reproduced via the memory-matrix formalism~\cite{Lucas:2015pxa}.\footnote{When implemented in concrete models such as the linear axion Gubser--Rocha model, however, this mechanism was found unable to capture both anomalous temperature dependences simultaneously~\cite{ahnInabilityLinearAxion2023}. Other holographic mechanisms proposed to address the same pair of scalings can be found in Refs.~\cite{Pal:2010sx,Kim:2010zq,Fadafan:2012hr,Zhou:2015dha,Ge:2016lyn,Blauvelt:2017koq,Cremonini:2018kla}.} The Gubser--Rocha model has also been applied to plasmons in layered strange metals~\cite{Eede:2023rrv}.

The classical-gravity approximation commonly used in applied holography is best controlled at large $N$ and strong coupling, while many standard constructions start from a UV conformal fixed point---conditions not fully realized in real materials. The semi-holographic framework~\cite{faulknerSemiHolographicFermiLiquids2011,faulknerIntegratingOutGeometry2011,mukhopadhyayPhenomenologicalCharacterisationSemiholographic2013,gursoyHolographyARPESSumrules2012} bridges this gap by coupling a weakly coupled UV sector to a strongly correlated holographic IR sector, retaining the predictive power for infrared physics while relaxing the requirements of pure holography. Recent studies have shown that semi-holography can reproduce $\rho \propto T$ over a wide temperature range and $\cot \theta_H \propto T^2$ above the Fermi temperature~\cite{doucotLinearin$T$ResistivitySemiholographic2021,Samanta:2022myh,doucotEffectiveFrameworkStrange2024}.

Of particular relevance to this work, the high-precision angle-resolved photoemission spectroscopy (ARPES) measurements by Smit et al.~\cite{smitMomentumdependentScalingExponents2024} revealed non-Lorentzian nodal spectral lineshapes in the cuprate
$\mathrm{(Pb,Bi)_2Sr_{2-x}La_xCuO_{6+\delta}}$
that cannot be captured by the momentum-independent power-law liquid (PLL) model~\cite{reberUnifiedFormLowenergy2019}, a framework that provides a unified description of the nodal self-energy across the entire cuprate doping phase diagram with a single smoothly varying exponent $\alpha$. The observed spectral asymmetry is well reproduced by promoting the PLL scaling exponent to a momentum-dependent function, a modification motivated by the Gubser--Rocha model within the semi-holographic framework. Building on this insight, Mauri et al.~\cite{mauriGaugegravityDualityComes2024} constructed a semi-holographic model that captures the lineshape asymmetry across doping levels at low temperatures but underestimates the thermal contribution to the self-energy at elevated temperatures. They identified three possible directions: (i) a more refined dual gravitational theory within the conformal-to-$\mathrm{AdS}_2$ class, (ii) a temperature-dependent coupling $g_k^2(T)$, and (iii) additional contributions to $G_0(\omega)$ beyond the electron--phonon term. In this paper, we focus on the first direction.

Testing the geometric possibility is naturally a holographic inverse problem: rather than selecting a bulk theory and computing its boundary response, one asks which bulk profiles are compatible with a given fermionic spectrum. More broadly, reconstructing bulk physics from boundary data is important both for building bottom-up models and for understanding spacetime emergence and quantum gravity~\cite{harlowTASILecturesEmergence2018,dejonckheereModaveLecturesBulk2018}. Various methods on this respect include smearing-function reconstruction for bulk operators~\cite{hamiltonLocalBulkOperators2006,hamiltonHolographicRepresentationLocal2006}, geodesic probes via boundary two-point functions~\cite{balasubramanianHolographicParticleDetection2000,hubenyExtremalSurfacesBulk2012}, inverse scattering from scalar and vector correlators~\cite{fanInverseProblemCorrelation2024,Fan:2025fxt}, and the derivation of the (1+1)d geometry using generalized free fields but without assuming symmetry matching~\cite{Nebabu:2026lmn,Nebabu:2023iox}. Quantum information-theoretic approaches have also proven fruitful, including entanglement wedge reconstruction~\cite{dongReconstructionBulkOperators2016,faulknerBulkLocalityModular2017,espindolaEntanglementWedgeReconstruction2018,peningtonEntanglementWedgeReconstruction2020}, holographic entanglement entropy~\cite{Hammersley:2007ab,Bilson:2008ab,Bilson:2010ff,jokelaPrecisionHolography2021,Jokela:2025ime,Ji:2025vks}, complexity~\cite{Hashimoto:2021umd,Xu:2023eof}, and tensor networks~\cite{Qi:2013caa,pastawskiHolographicQuantumErrorcorrecting2015,caoBuildingBulkGeometry2020,Bao:2025plr,Geng:2025efs}. Pole-skipping points of boundary Green's functions provide yet another window into the near-horizon geometry~\cite{luBulkSpacetimeEncoding2026,luAlgebraicStructureUnderlying2026,Ran:2026uog}. While these methods have significantly advanced our understanding of holographic encoding, they often rely on specific assumptions about boundary observables or the bulk geometry.

Machine learning has opened new avenues for holographic inverse problems. Hashimoto et al.~\cite{hashimotoDeepLearningAdS2018} established the AdS/DL correspondence by interpreting the radial coordinate as network depth. Neural ODEs~\cite{chenNeuralOrdinaryDifferential2018}, which parameterize the derivative of a hidden state with a neural network, are naturally suited to holographic radial evolution, while Physics-Informed Neural Networks (PINNs)~\cite{raissiPhysicsinformedNeuralNetworks2019} complement this approach by incorporating equations of motion into the loss function. These techniques have been applied to reconstruct bulk geometries from QCD observables~\cite{hashimotoDeepLearningHolographic2018,akutagawaDeepLearningAdS2020,Hashimoto:2020jug,Luo:2024iwf,Filev:2025mbt,ThomasArun:2025uyi,Zhu:2026nnqcd}, shear viscosity~\cite{yanDeepLearningBlack2020,guNeuralODEsHolographic2025}, equations of state~\cite{chenMachineLearningHolographic2024,Bea:2024xgv}, optical conductivity~\cite{liLearningBlackHole2023,ahnDeepLearningBulk2024,ahnDeepLearningbasedHolography2025}, and entanglement entropy~\cite{ahnHolographicReconstructionBlack2025,Hashimoto:2025zmi}. In particular, Ref.~\cite{Ran:2025vat} combined a scalar Neural-ODE inverse problem with the holographic Wilsonian renormalization group (HWRG). Its construction is semi-holographic in a Wilsonian sense: a finite radial cutoff retains the dynamical IR bulk scalar while the integrated UV geometry is encoded in a boundary action containing source and double-trace terms. This framework was used to learn geometries with AdS and non-AdS asymptotics from the normalized scalar correlator with frequency dependence.

The strange-metal application considered here uses a different semi-holographic construction, following Refs.~\cite{smitMomentumdependentScalingExponents2024,mauriGaugegravityDualityComes2024}: a weakly coupled boundary electron interacts through $g_k$ with a fermionic operator of a strongly coupled IR sector, whose holographic Green's function supplies the electronic self-energy. Thus, the two settings share an IR holographic sector but implement the UV--IR matching differently. We adopt the inverse-learning strategy of Ref.~\cite{Ran:2025vat}, while the physical semi-holographic decomposition and the target observable follow Refs.~\cite{smitMomentumdependentScalingExponents2024,mauriGaugegravityDualityComes2024}. The resulting finite-momentum fermionic problem has a distinct source--response structure and spinorial in-falling condition, and we jointly learn effective metric profiles and the charge-weighted gauge potential $qA_t$. The importance of the fermionic inverse problem is evident. As emphasized in Chapter~9 of Ref.~\cite{zaanenHolographicDualityCondensed2015}, the single-fermion two-point function, which can be probed by ARPES, is a uniquely powerful diagnostic of strongly correlated electron matter: unlike the collective bosonic responses that underlie previous holographic machine-learning studies, it carries the imprint of fermionic quantum statistics, resolves the $(\omega,k)$ dependence over a broad kinematic range rather than only the small-momentum hydrodynamic regime, and directly registers the Fermi-surface structure of the finite-density ground state.

We first benchmark the fermionic reconstruction on the analytically tractable Einstein--Maxwell and Gubser--Rocha models. We then apply the framework to nodal strange-metal phenomenology in a semi-holographic setting. We generate the training targets analytically from the normalized extended-PLL model with ARPES-calibrated parameters. This construction isolates the geometric direction identified above:\footnote{Because we use the normalized extended-PLL interaction contribution as the target, the frequency-independent factor $g_k^2(T)$ cancels at fixed $(k,T)$. Hence, if $G_0$ is unchanged, the observed tension between spectral-shape fitting accuracy and frequency symmetry cannot be remedied by direction (ii). This statement concerns the normalized spectral shape; $g_k^2(T)$ can still change the absolute linewidth.} within the prescribed conformal-to-$\mathrm{AdS}_2$ class, we search for effective bulk representatives beyond the fixed analytical Gubser--Rocha IR benchmark rather than fitting measured MDC intensities directly~\cite{mauriGaugegravityDualityComes2024,smitMomentumdependentScalingExponents2024}. For this application, we choose a massless effective Dirac probe, motivated jointly by the linear extended-PLL scaling exponent and boundary regularity, as explained in Section~\ref{sec:ansatz}. We adapt the conserved-Dirac-flux method of Ref.~\cite{GulottaHerzogKaminski2011} to the conformal-to-$\mathrm{AdS}_2\times\mathbb{R}^2$ asymptotics considered here. The resulting formula extracts the spectral function without isolating the subleading response coefficient. The same flux construction guarantees spectral positivity for every real output allowed by our metric and gauge-field parametrizations. We identify two analytic redundancies that must be accounted for before the raw learned profiles can be compared. After fixing the coordinate/Weyl redundancy and mapping temperature-degenerate solutions to a common scaled Hawking temperature, we assess the performance of the present framework across doping levels and temperatures.

The remainder of this paper is organized as follows. Section~\ref{sec:holo-fermions} reviews the standard procedure for computing fermionic Green's functions in holography. Section~\ref{sec:semi-holo} introduces the semi-holographic framework and the PLL model with momentum-dependent scaling exponents. Section~\ref{sec:ansatz} presents the bulk ansatz and derives the master equations for geometries with either $\mathrm{AdS}_4$ or conformal-to-$\mathrm{AdS}_2 \times \mathbb{R}^2$ asymptotics. Section~\ref{sec:ml} describes the Neural ODE architecture, loss function, and training protocol. Numerical results for the Einstein--Maxwell theory, Gubser--Rocha model, and the extended PLL targets are presented in Section~\ref{sec:examples}. Finally, Section~\ref{sec:conclusion} summarizes our findings and discusses their implications. Appendix~\ref{app:ir-green} details the derivation of the IR Green's function and effective geometry for the Gubser--Rocha model, Appendix~\ref{app:flux} derives the flux-based formula used to extract the spectral function in the geometry with conformal-to-$\mathrm{AdS}_2\times\mathbb{R}^2$ asymptotics, Appendix~\ref{app:F} fixes the residual gauge freedom by matching the frame of the standard $\mathrm{AdS}_2\times\mathbb{R}^2$ black hole, Appendix~\ref{app:units} discusses the explicit restoration of SI units to map bulk variables to experimental observables, Appendix~\ref{app:th_degeneracy} derives the temperature map and the resulting degeneracy of the normalized spectral function for the massless Dirac system in the constant-$h$ frame, and Appendix~\ref{app:robustness} collects numerical validation tests and diagnostics for the extended-PLL fits.

\section{Holographic Fermionic Green's Function}\label{sec:holo-fermions}

In this section, we briefly review the standard holographic procedure for computing the retarded Green's function of a fermionic operator via the AdS/CFT correspondence. In this setup, the bulk metric and gauge potential are the inputs and the boundary correlation functions are the outputs.

Consider a four-dimensional static black hole with planar symmetry, whose line element takes the form%
\begin{equation}
ds^{2}=g_{tt}(r)dt^{2}+g_{rr}(r)dr^{2}+g_{xx}(r)(dx^{2}+dy^{2}).
\end{equation}%
A Dirac spinor field $\psi $ of mass $m$ and charge $q$ on this background obeys the curved-spacetime Dirac equation%
\begin{equation}
\left( \Gamma ^{\mu}D_{\mu}-m\right) \psi =0,
\end{equation}%
where $\Gamma^\mu=e^\mu{}_{a}\Gamma^a$, with $e^\mu{}_{a}$ the inverse vielbein and $\Gamma^a$ the tangent-space gamma matrices. The covariant derivative is $D_{\mu}=\partial _{\mu}+\frac{1}{4}%
\omega _{\mu ab}\Gamma ^{ab}-iqA_{\mu}$, with $\omega _{\mu ab}$ the spin connection, $\Gamma ^{ab}\equiv \frac{1}{2}[\Gamma ^a,\Gamma ^b]$, and $A_{\mu}$ the $U(1)$ gauge field. The fermion mass is typically restricted to the range $|m|<1/2$, within which both standard and alternative quantizations of the dual boundary operator are admissible.
To solve the Dirac equation, we perform a Fourier decomposition and introduce the rescaled spinor components
\begin{equation}
\psi _{\pm }=(-gg^{rr})^{-1/4}e^{-i\omega t+ikx}\phi _{\pm },
\end{equation}%
which reduces the Dirac equation to a coupled system of first-order ODEs for $\phi _{\pm }$:
\begin{equation}
\sqrt{\frac{g_{xx}}{g_{rr}}}(\partial _{r}\mp m\sqrt{g_{rr}})\phi _{\pm
}=\pm i(w\gamma ^{0}-k\gamma ^{1})\phi _{\mp },
\end{equation}%
where%
\begin{equation}
w=\sqrt{\frac{g_{xx}}{-g_{tt}}}(\omega +qA_{t}).
\end{equation}

Computing the retarded two-point function of the boundary fermionic operator $\mathcal{O}$ dual to the bulk Dirac field requires solving this system subject to appropriate boundary conditions. At the black hole horizon, one imposes purely in-falling boundary conditions. Near the asymptotic AdS boundary, $\phi _{\pm }$ exhibit the power-law behavior
\begin{equation}
\phi _{+}\approx A(\omega ,k)r^{m}+B(\omega ,k)r^{-m-1},\;\phi _{-}\approx C(\omega ,k)r^{m-1}+D(\omega ,k)r^{-m} .  \label{faiA}
\end{equation}%
The retarded Green's function is then given by the holographic prescription~\cite{iqbalRealtimeResponseAdS2009}%
\begin{equation}
G_{R}(\omega ,k)=-i\mathcal{S}(\omega ,k)\gamma ^{0},
\end{equation}%
where the matrix $\mathcal{S}(\omega ,k)$ encodes the linear relation between the source and response coefficients,
\begin{equation}
D(\omega ,k)=\mathcal{S}(\omega ,k)A(\omega ,k).
\end{equation}

With the gamma-matrix basis $\gamma ^{0}=i\sigma ^{2}$, $\gamma ^{1}=\sigma^{1}$, $\gamma ^{2}=\sigma ^{3}$ and the decomposition $\phi _{\pm }=(-iy_{\pm },z_{\pm })^{T}$ of each two-component spinor, the system decouples into two independent sets of equations:
\begin{align}
\sqrt{\frac{g_{xx}}{g_{rr}}}(\partial _{r}\mp m\sqrt{g_{rr}})y_{\pm }& =\pm
(k-w)z_{\mp },  \label{yz1} \\
\sqrt{\frac{g_{xx}}{g_{rr}}}(\partial _{r}\pm m\sqrt{g_{rr}})z_{\mp }& =\pm
(k+w)y_{\pm }.
\label{yz2}
\end{align}%
Introducing the ratios%
\begin{equation}
\xi _{+}=\frac{y_{-}}{z_{+}},\quad \xi _{-}=\frac{z_{-}}{y_{+}},
\label{Xizf}
\end{equation}%
Eqs.~(\ref{yz1}) and~(\ref{yz2}) yield two decoupled radial flow equations:
\begin{equation}
\sqrt{\frac{g_{xx}}{g_{rr}}}\partial _{r}\xi _{\pm }=-2m\sqrt{g_{xx}}\xi
_{\pm }\mp (k\mp w)\pm (k\pm w)\xi _{\pm }^{2}.
\label{floweq}
\end{equation}%
The in-falling boundary condition at the horizon $r_{h}$ then reduces to the constant initial value\footnote{%
The in-falling boundary condition is different exactly at $\omega = 0$~\cite{liuNonFermiLiquidsHolography2011}. To avoid the complexity, we assume $\omega \neq 0$ throughout this paper.}%
\begin{equation}
\xi _{\pm }|_{r=r_{h}}=i,
\end{equation}%
and the retarded Green's function is extracted at the boundary as%
\begin{equation}
G=\lim_{\epsilon \rightarrow 0}\epsilon ^{-2m}%
\begin{pmatrix}
\xi _{+} & 0 \\ 
0 & \xi _{-}%
\end{pmatrix}%
\bigg|_{r=\frac{1}{\epsilon }}.
\label{GX}
\end{equation}

Finally, the flow equations for $\xi _{\pm }$ are mapped into each other under $k\rightarrow -k$, which implies the symmetry relation
\begin{equation}
G_{11}(\omega ,-k)=G_{22}(\omega ,k).
\end{equation}%
Without loss of generality, we therefore restrict our attention to the $\xi_{-}$ branch of the flow equation~(\ref{floweq}) and to the spectral function determined by the imaginary part of $G_{22}$.

It is worth noting that this prescription manifestly respects spectral positivity. For a real background and real $\omega\neq0$ and $k$, write $\xi_-=a+ib$. The imaginary part of the flow equation~(\ref{floweq}) is homogeneous in $b$,
\begin{equation}
\sqrt{\frac{g_{xx}}{g_{rr}}}\,b'=-2\left[m\sqrt{g_{xx}}+(k-w)\,a\right]b,
\end{equation}
so $b$ never changes sign along the radial flow. The in-falling condition $\xi_-|_{r=r_h}=i$ fixes $b(r_h)=1>0$, hence $b(r)>0$ everywhere. Since the boundary prefactor in Eq.~(\ref{GX}) is positive, $\mathrm{Im}\,G_{22}=\lim_{\epsilon\rightarrow0}\epsilon^{-2m}\,b|_{r=1/\epsilon}>0$. This holds for any real metric and gauge profile, not only for solutions of the equations of motion.
\section{Semi-Holographic Power-Law Liquid}\label{sec:semi-holo}

The purely holographic calculation reviewed above yields the Green's function of a composite operator, whereas ARPES measures the spectral function of the physical electron. We therefore combine the phenomenological PLL model for strange metals~\cite{reberUnifiedFormLowenergy2019} with the semi-holographic framework, which couples these degrees of freedom and provides the basis for the momentum-dependent extension of the PLL model~\cite{smitMomentumdependentScalingExponents2024,mauriGaugegravityDualityComes2024}.

\subsection{Power-Law Liquid Model}\label{subsec:pll}

To capture the anomalous scaling of the imaginary part of the electronic self-energy in the normal state of cuprates, Reber et al.~\cite{reberUnifiedFormLowenergy2019} introduced the PLL model. Based on detailed nodal ARPES measurements of $\text{Bi-2212}$, this phenomenological model provides a unified description of non-Fermi liquid interactions over a wide doping range, interpolating between the marginal Fermi liquid~\cite{varmaPhenomenologyNormalState1989} and the hidden Fermi liquid~\cite{caseyHiddenFermiLiquid2011}.

In ARPES experiments, under the assumption of a momentum-independent self-energy, the momentum distribution curves (MDCs) at fixed binding energy $\hbar\omega$ are well described by a Lorentzian profile,
\begin{equation}
\mathcal{A}(\omega ,k)=\frac{W}{\pi }\frac{\Gamma /2}{(k-k_{\ast
})^{2}+(\Gamma /2)^{2}},
\end{equation}%
where $\mathcal{A}(\omega,k)$ is the single-particle spectral function measured by ARPES, $W$ denotes its intensity, $k_{\ast }$ the peak position, and $\Gamma$ the MDC full width at half maximum. Following Refs.~\cite{reberUnifiedFormLowenergy2019,mauriGaugegravityDualityComes2024}, we write $\Sigma^{R}=\Sigma^{\prime}-i\Sigma^{\prime\prime}$ with $\Sigma^{\prime\prime}>0$. The quantity $\Sigma^{\prime\prime}$ is the negative of the imaginary part of the retarded electronic self-energy extracted from ARPES; for a linearized dispersion it is related to the MDC width by $\Gamma=2\Sigma^{\prime\prime}/v_F$, where $v_F$ is the renormalized Fermi velocity. The PLL form is then written as
\begin{align}
\Sigma_{\mathrm{PLL}}^{\prime \prime }(\omega,T)
&=G_{0}(\omega,T)+\Sigma_{\mathrm{int}}^{\prime \prime }(\omega,T),
\label{selfenergy}
\\
\Sigma_{\mathrm{int}}^{\prime \prime }(\omega,T)
&=\lambda \frac{\lbrack (\hbar \omega )^{2}+(\beta k_{B}T)^{2}]^{\alpha }}{%
(\hbar \omega _{N})^{2\alpha -1}},
\label{sigmaPLL}
\end{align}%
where $G_{0}(\omega,T)$ denotes the additive self-energy contribution from impurity scattering and electron--phonon coupling~\cite{smitMomentumdependentScalingExponents2024,mauriGaugegravityDualityComes2024}. The model is parameterized by a dimensionless coupling constant $\lambda \sim 0.5$ characteristic of Planckian dissipation, a temperature scaling factor $\beta \sim \pi$, and a normalization scale $\hbar\omega_N{}=0.5\text{ eV}$. Crucially, the scaling exponent $\alpha$ encodes the nature of the electronic interactions: it evolves smoothly from a quadratic Fermi liquid ($\alpha=1$) in the overdoped regime, through a linear marginal Fermi liquid ($\alpha=1/2$) at optimal doping, to a non-Fermi liquid in the underdoped phase.

Although the isotropic PLL model successfully captures the $\omega/T$ scaling of the nodal self-energy over a wide parameter range, recent high-precision MDC measurements by Smit et al.~\cite{smitMomentumdependentScalingExponents2024} have revealed an intrinsic asymmetry away from the Fermi level, with spectral weight shifting toward $|k|>|k_{\ast }|$. This asymmetry signals a finite momentum dependence of the self-energy. To accommodate it, an extended PLL model was proposed~\cite{smitMomentumdependentScalingExponents2024} in which the scaling exponent acquires a $k$-dependence:
\begin{align}
\Sigma _{\mathrm{E}}^{\prime \prime }(\omega,k)
&=\lambda \frac{\lbrack (\hbar \omega )^{2}+(\beta k_{B}T)^{2}]^{\alpha
\left( k\right) }}{(\hbar \omega _{N})^{2\alpha \left( k\right) -1}},
\label{sigmaHpll} \\
\alpha \left( k\right) &=\alpha \left[ 1-\left( \frac{k-k_{F}}{k_{F}}%
\right) \right] .
\label{ak}
\end{align}%
Here $\Sigma _{\mathrm{E}}^{\prime \prime }$ denotes the momentum-dependent interaction contribution to the electronic self-energy, and $k_{F}$ is the Fermi momentum. Equation~\eqref{ak} adopts the Gubser--Rocha value $V=-1$ in the parametrization $\alpha(k)=\alpha[1+V(k-k_F)/k_F]$. This choice captures the measured MDC asymmetry very well, although fits with free $V$ show modest frequency-dependent departures from $-1$~\cite{smitMomentumdependentScalingExponents2024}. We keep $V=-1$ fixed throughout this work.
\subsection{Semi-Holographic Coupling}\label{subsec:semi-holo-framework}

The semi-holographic framework~\cite{faulknerSemiHolographicFermiLiquids2011,gursoyHolographyARPESSumrules2012,mauriGaugegravityDualityComes2024} couples a weakly coupled UV electron sector, described by the field $\Psi$, to a strongly coupled IR sector represented by a composite operator $\mathcal{O}$. The former describes the physical electron whose spectral function is measured by ARPES, while the latter is governed by holographic duality and encodes the quantum-critical behavior.

The minimal effective action coupling these two sectors takes the form
\begin{equation}
S_{\text{eff}}=\int \frac{d\omega d^{2}k}{(2\pi )^{3}}\left[ \Psi ^{\dagger
}(-\hbar \omega +\epsilon (k)-\mu )\Psi +g_{k}\Psi ^{\dagger }\mathcal{O}%
+g_{k}\mathcal{O}^{\dagger }\Psi \right] +S_{\text{strong}}(\mathcal{O}),
\end{equation}%
where $g_{k}$ is a coupling constant assumed to be real and momentum-dependent~\cite{mauriGaugegravityDualityComes2024}. Using this action, the full retarded Green's function of the physical electron is given by
\begin{equation}
G_{\Psi
\Psi }(\omega ,k)=\frac{\hbar }{-\hbar \omega +\epsilon (k)-\mu +\Sigma(\omega,k)},
\end{equation}%
from which the electron self-energy reads as
\begin{equation}
\Sigma (\omega ,k)=-g_{k}^{2}\mathcal{G}_{k}(\omega).
\label{sigma}
\end{equation}%
For comparison with MDC data near the Fermi surface, Ref.~\cite{mauriGaugegravityDualityComes2024} further linearizes the dispersion and absorbs the real part of the self-energy into the renormalized velocity $v_F$. The resulting electron Green's function takes the form
\begin{equation}
G_{\Psi\Psi}^{\mathrm{MDC}}(\omega,k)
=\frac{\hbar}{-\hbar\omega+ v_F(k-k_F)-i g_k^2\,\mathrm{Im}\,\mathcal{G}_k(\omega)}.
\label{eq:semi-holo-mdc}
\end{equation}%
Equation~\eqref{eq:semi-holo-mdc} makes explicit the central observation used below: the imaginary part of the self-energy, $\Sigma^{\prime \prime }(\omega ,k)$, which governs the lifetime of electronic excitations and shapes the ARPES lineshape, is controlled by $g_k^2\,\mathrm{Im}\,\mathcal{G}_{k}(\omega)$.

\subsection{Holographic IR Green's Function}\label{subsec:grav-contribution}

In Ref.~\cite{mauriGaugegravityDualityComes2024}, the semi-holographic framework was used to provide a theoretical foundation for the momentum dependence of the self-energy by identifying $\mathcal{G}_k$ with the IR Green's function of the Gubser--Rocha model---an EMD theory in $(3+1)$-dimensional spacetime. The charged black hole solution of this model is asymptotically AdS$_4$, with an interior geometry rendered nontrivial by the dilaton. In the low-temperature, low-energy limit of the Dirac equation, both the mass term ($\sim m$) and the gauge term ($\sim qA_{t}$) become subleading at small $r/\mu $, and the resulting equation can be solved analytically. This suppression occurs in the IR scaling limit and does not require the original bulk fermion mass to vanish. A detailed review of this derivation is given in Appendix~\ref{app:ir-green}. The key result is that the effective IR geometry has conformal-to-AdS$_{2}\times \mathbb{R}^{2}$ asymptotics:%
\begin{equation}
ds^{2}=\Omega(z) \left[ \frac{-dt^{2}+dz^{2}}{z^{2}}+h_{0}(dx^{2}+dy^{2})%
\right] ,  \label{adsr2}
\end{equation}%
where $z\sim r^{-1/2}$, $h_{0}$ is a constant, and $\Omega(z)$ is a conformal factor. The Dirac equation in this effective geometry admits, on the branch relevant to our application, the asymptotic behavior
\begin{equation}
\left( 
\begin{array}{c}
y_{+} \\ 
z_{-}%
\end{array}%
\right) =\left( 
\begin{array}{c}
-1 \\ 
1%
\end{array}%
\right) \left( R+J_{\pm }z\right) z^{\nu_{k}}+\left( 
\begin{array}{c}
1 \\ 
1%
\end{array}%
\right) \left( S+K_{\pm }z\right) z^{-\nu_{k}}+\cdots.
\label{yzA}
\end{equation}%
Crucially, the exponent is $k$-dependent: $\nu_{k}=\lvert k\rvert/\sqrt{h_{0}}$. As shown in Ref.~\cite{mauriGaugegravityDualityComes2024}, the bulk holographic fermion is dual to a composite operator describing hole excitations. Since ARPES measures the electronic response, a particle-hole conjugation $k \to 2k_F - k$ is required (see Section~\ref{subsec:pll-k} for details). This transforms the effective scaling exponent into $\nu_{2k_F-k}=\lvert2k_F-k\rvert/\sqrt{h_0}$. Throughout our sampled window, $2k_F-k>0$, so the absolute value can be omitted when matching this exponent to the phenomenological form. By defining $\alpha \equiv \nu_{k_F}$ at the Fermi surface, one then exactly recovers the linear momentum dependence $\alpha(k) = \alpha[1-(k-k_F)/k_F]$ postulated in Eq.~(\ref{ak})~\cite{smitMomentumdependentScalingExponents2024}. Furthermore, treating the coefficients $R$ and $S$ as the response and source, respectively, the IR Green's function of Ref.~\cite{mauriGaugegravityDualityComes2024} is given by their ratio\footnote{We have suppressed an $\omega$-independent prefactor. In the self-energy this factor can be absorbed into $g_{k}^{2}$, and it will not affect our training data in Section \ref{subsec:pll-k}.}
\begin{equation}
\mathcal{G}_{k}\sim \frac{R}{S}.
\label{RS}
\end{equation}%

One issue that must be addressed is the numerical extraction of $\mathcal{G}_{k}$. Unlike the asymptotic behavior~(\ref{faiA}) in AdS$_{4}$, where the relevant coefficients appear at leading order, the response coefficient $R$ in a geometry conformal to AdS$_{2}\times \mathbb{R}^{2}$ is subleading for $\nu_{k}>0$, and the standard extraction formula~(\ref{GX}) no longer applies.

For our purposes, this issue can be resolved without determining $R$ separately, because the training target depends only on $\mathrm{Im}\,\mathcal{G}_{k}$. We use the conserved-Dirac-flux method of Ref.~\cite{GulottaHerzogKaminski2011}, identifying the source projection and normalization appropriate to the present conformal-to-$\mathrm{AdS}_2\times\mathbb{R}^2$ asymptotics. For the massless effective probe chosen below, with real background fields and real $\omega$ and $k$, the conserved radial flux $\mathcal{J}=\mathrm{Im}(y_{+}\bar{z}_{-})$ then yields the required imaginary part:
\begin{equation}
\mathrm{Im}\,\mathcal{G}_{k}=\mathrm{Im}\frac{R}{S}
=-2\mathcal{J}\left.
\frac{z^{-2\nu_{k}}}{\left| y_{+}+z_{-}\right| ^{2}}\right\vert
_{z\rightarrow 0}.  \label{GXi}
\end{equation}%
Here $y_++z_-$ is the source projection for the branch relevant to our application. See the derivation in Appendix~\ref{app:flux}.

This flux-based expression can be evaluated directly from the solution $(y_{+},z_{-})$ and is insensitive to its overall normalization.

Before applying this framework to numerical data, we must first incorporate certain physical priors and parameterize the bulk fields. The next section develops the bulk-field formulation required for this purpose.

\section{Bulk Ansatz and Master Equations}\label{sec:ansatz}

Building on the holographic and semi-holographic frameworks established above, we specify two distinct ansätze for the bulk fields and derive their associated master equations, which form the basis of our Neural ODE algorithm.

The first ansatz parametrizes a static, planar, asymptotically AdS$_4$ black-hole background for the probe Dirac equation. The line element reads%
\begin{equation}
ds^{2}=\frac{1}{z^{2}}\left[ -f(z)dt^{2}+\frac{1}{f(z)}%
dz^{2}+h(z)(dx^{2}+dy^{2})\right] ,  \label{ads4}
\end{equation}%
where the two metric functions are parametrized as%
\begin{equation}
f(z)=(1-z)e^{n_f(z)z},\;h(z)=e^{n_{h}(z)}.
\label{fh}
\end{equation}%
The charge-weighted gauge potential is parametrized as
\begin{equation}
qA_{t}(z)=(1-z)n_{a}(z)^{2}.
\label{eq:qA_ads4}
\end{equation}%
Here $n_f(z)$, $n_h(z)$, and $n_a(z)$ are undetermined functions to be represented by independent neural networks in Section~\ref{sec:ml}. By construction, the horizon lies at $z=1$ and the boundary at $z=0$, with $f/(1-z)$ and $h$ positive and $qA_t/(1-z)\ge0$.\footnote{This sign-definite prior is motivated by the usual Maxwell and EMD models; sign-changing profiles would require a more general effective electromagnetic sector beyond the present prior.} Since $qA_{t}$ enters the Dirac equation as a single combination, we parametrize it by one neural network rather than modeling $q$ and $A_{t}$ separately.

Substituting the line element~(\ref{ads4}) into the radial flow equation~(\ref{floweq}) yields%
\begin{equation}
\frac{w+k}{\sqrt{fh}}-\frac{2m\xi _{-}}{z\sqrt{f}}+\frac{w-k}{\sqrt{fh}}\xi
_{-}^{2}+\xi _{-}^{\prime }=0,  \label{Hflow}
\end{equation}%
where%
\begin{equation}
w=\sqrt{\frac{h}{f}}\left( \omega +qA_{t}\right) ,
\end{equation}%
and we have focused on the $\xi _{-}$ branch~\cite{liuNonFermiLiquidsHolography2011}.

The second ansatz is a finite-temperature extension of the conformal-to-AdS$_2 \times \mathbb{R}^2$ geometry in Eq.~(\ref{adsr2}):%
\begin{equation}
ds^{2}=\Omega(z)\left\{ \frac{1}{z^{2}}\left[ -f(z)dt^{2}+\frac{dz^{2}}{f(z)}%
\right] +h(z)(dx^{2}+dy^{2})\right\} .  \label{cadsr2}
\end{equation}%
Here the two metric functions ($f$ and $h$) and the charge-weighted gauge potential $qA_t$ are parametrized as in the first ansatz, Eqs.~(\ref{fh}) and~(\ref{eq:qA_ads4}); the conformal factor $\Omega(z)$ is kept explicit at this stage.
Substituting the line element~(\ref{cadsr2}) into the Dirac equations~(\ref{yz1}) and~(\ref{yz2}) yields the coupled master equations%
\begin{equation}
\begin{aligned}
y_{+}' + \frac{m\sqrt{\Omega}}{z\sqrt{f}}y_+ - \frac{w-k}{z\sqrt{fh}}z_- &= 0, \\
z_-' - \frac{m\sqrt{\Omega}}{z\sqrt{f}}z_- + \frac{w+k}{z\sqrt{fh}}y_+ &= 0,
\end{aligned} \label{yz34}
\end{equation}%
where%
\begin{equation}
w=z\sqrt{\frac{h}{f}}\left( \omega +qA_{t}\right) .
\end{equation}%
The master equations enjoy two scaling symmetries,
\begin{align}
\omega &\rightarrow s\omega ,\;k\rightarrow sk,\;f\rightarrow
sf,\;h\rightarrow sh,\;A_{t}\rightarrow sA_{t},\;m\rightarrow s^{1/2} m,  \label{ss1} \\
k &\rightarrow sk,\;h\rightarrow s^{2}h.
\label{ss2}
\end{align}%
The first is broken by our ansatz, which fixes $f(0)=1$ for any neural network $n_{f}(z)$ with finite output on $[0,1]$. The second will be exploited to set the unit of momentum.

To solve the master equations~(\ref{yz34}), we impose the in-falling boundary condition on $(y_+,z_-)$ in the near-horizon region, which fixes the relation $iy_+=z_-$. Because the Dirac equation is linear and the extraction formula~(\ref{GXi}) is insensitive to the overall normalization, we adopt $(y_+,z_-)=(1,i)$ for convenience.

In Section~\ref{subsec:pll-k}, we solve Eq.~(\ref{yz34}) with this in-falling boundary condition and then extract the spectral function $\mathrm{Im}\,\mathcal{G}_{k}$ via Eq.~(\ref{GXi}). The procedure, however, requires prior knowledge of the scaling exponent $\nu_{k}$. We now determine $\nu_{k}$ in general, following Chapter 9 of Ref.~\cite{zaanenHolographicDualityCondensed2015} for the RN-AdS black hole. The coupled master equations~(\ref{yz34}) can be written in matrix form as%
\begin{equation}
\partial _{z}\left( 
\begin{array}{c}
y_{+} \\ 
z_{-}%
\end{array}%
\right) +U\left( 
\begin{array}{c}
y_{+} \\ 
z_{-}%
\end{array}%
\right) =0,  \label{eigen1}
\end{equation}%
where the matrix $U$ is%
\begin{equation}
U=\left( 
\begin{array}{cc}
\frac{m\sqrt{\Omega }}{z\sqrt{f}} & -\frac{w-k}{z\sqrt{fh}} \\
\frac{w+k}{z\sqrt{fh}} & -\frac{m\sqrt{\Omega }}{z\sqrt{f}}%
\end{array}%
\right) .
\end{equation}%
Near the boundary, we assume the asymptotic behavior%
\begin{equation}
w\rightarrow \mathcal{O}\left( z\right) ,\;f\rightarrow \mathcal{O}\left(
1\right) ,\;h\rightarrow \mathcal{O}\left(
1\right) ,\;\left( 
\begin{array}{c}
y_{+} \\ 
z_{-}%
\end{array}%
\right) =z^{-\nu}\phi ,
\end{equation}%
where $\phi $ is a $z$-independent spinor.
Substituting into Eq.~(\ref{eigen1}) yields the eigenvalue equation%
\begin{equation}
\left( \left. zU\right\vert _{z\rightarrow 0}\right) \phi =\nu\phi .
\end{equation}%
Solving this eigenvalue problem yields the scaling exponents
\begin{equation}
\nu_{k}=\left.  \sqrt{\frac{k^2}{fh} + \frac{m^2\Omega}{f}}\right\vert _{z\rightarrow 0}.
\label{vk}
\end{equation}

Equation~(\ref{vk}) uses the bounded boundary potential imposed by Eq.~(\ref{eq:qA_ads4}), which gives $w=\mathcal{O}(z)$. This boundary regularity is a physical prior of the present ansatz and excludes more singular electric backgrounds (such as $A_t\sim z^{-1}$), for which the gauge field contributes to $\nu_k$. Within this prior, the mass term is the only source of nonlinear dependence of $\nu_k$ on the momentum $k$. In the Gubser--Rocha model that motivates the extended PLL framework, both the mass and the gauge terms become subleading in the near-horizon region~\cite{mauriGaugegravityDualityComes2024} (see also Appendix~\ref{app:ir-green}), so the IR scaling exponent naturally reduces to $\nu_k \propto \lvert k\rvert$. This linear momentum dependence is precisely the structure underlying the phenomenological exponent $\alpha(k)$ in Eq.~(\ref{ak}). Within our ansatz, reproducing this exact linear dependence requires the asymptotic mass contribution in Eq.~(\ref{vk}) to vanish. For a nonzero constant probe mass, this would require $\Omega(z)\to0$ as $z\to0$, since $f(0)=1$. Under the smooth asymptotic conditions of our ansatz, however, such a vanishing conformal factor leads to divergent curvature invariants.\footnote{Writing $g_{\mu\nu}=\Omega\bar g_{\mu\nu}$, the Weyl invariant obeys $C_{\mu\nu\rho\sigma}C^{\mu\nu\rho\sigma}=\Omega^{-2}\bar C_{\mu\nu\rho\sigma}\bar C^{\mu\nu\rho\sigma}\sim1/\Omega^2$ for the smooth $\mathrm{AdS}_2\times\mathbb{R}^2$ asymptotics of $\bar g$.} We therefore choose $m=0$ for the effective probe model, reproducing the target scaling without requiring this singular boundary behavior.

With the effective-model choice $m=0$, the conformal factor drops out of the full master equations~(\ref{yz34}), so the fermionic spectra constrain only the conformal class of the bulk geometry. The pair $\{f(z),h(z)\}$ is therefore not separately physical: a radial reparametrization $z\to u(z)$ with a compensating Weyl rescaling can trade content between $f$ and $h$ while leaving the fermionic observable invariant. We keep both as trainable networks for optimization reasons\footnote{Empirically, forcing a single metric function to absorb all geometric deformations can result in relatively more complex profiles and a slightly higher training loss.}, but the spectroscopically meaningful content they carry is gauge invariant. To exhibit this content and to compare with the standard $\mathrm{AdS}_2\times\mathbb{R}^2$ black hole---whose spatial metric is constant---we use the gauge freedom to fix the spatial metric to its constant boundary value $h_0=h(0)$. As shown in Appendix~\ref{app:F}, in this frame the geometry is captured by $h_0$ and the blackening factor $F(u)$, defined in Eq.~(\ref{eq:Fdef}).

\section{Neural ODE Algorithm}\label{sec:ml}

\subsection{Neural ODE Architecture}\label{subsec:neural-ode}

Following the general Neural-ODE formulation used in Ref.~\cite{Ran:2025vat}, we represent radial evolution as a continuous-depth network. The update rule of a residual network, $x_{t+1}=x_t+Y(x_t)$, can be viewed as a forward-Euler discretization of an ordinary differential equation; Neural ODEs~\cite{chenNeuralOrdinaryDifferential2018} instead parameterize the derivative of the hidden state $x(t)$ by a neural network $y$,%
\begin{equation}
\frac{dx(t)}{dt}=y(x(t),t,\theta ).
\end{equation}%
with the output obtained by integrating from $t_0$ to $t_1$,
\begin{equation}
x(t_{1})=x(t_{0})+\int_{t_{0}}^{t_{1}}y(x(t),t,\theta )dt.
\end{equation}%
The continuous radial coordinate makes this architecture natural for holographic flow equations, although repeated ODE solves make the training computationally demanding. The substantive extension relative to the scalar, zero-momentum flow in Ref.~\cite{Ran:2025vat} lies in the hidden state and the trainable bulk content. We separate the complex Dirac equations into real and imaginary parts, identify $z$ with the Neural-ODE ``time,'' and use either the ratio $\xi_-$ or the coupled components $(y_+,z_-)$ as the state. The data span a two-dimensional $(\omega,k)$ grid, and three independent networks jointly represent $\{n_f(z),n_h(z),n_a(z)\}$, which encode two metric functions and $qA_t$. Each is a multilayer perceptron with three dense layers, $1\to10$, $10\to10$ with $\tanh$ activation, and $10\to1$, giving 141 trainable parameters initialized from a standard fan-in uniform distribution.
\subsection{Loss Function}\label{subsec:loss}

Throughout this work, we adopt a single grid-averaged loss function for all tasks,
\begin{equation}
L=\frac{1}{2N_{\mathcal{D}}}\sum_{(\omega,k)\in\mathcal{D}}\left(
\frac{\chi _{\mathrm{NODE}}(\omega,k)}{\chi _{\mathrm{DATA}}(\omega,k)}
-1-\ln \frac{\chi _{\mathrm{NODE}}(\omega,k)}{\chi _{\mathrm{DATA}}(\omega,k)}
\right) ,
\label{eq:loss}
\end{equation}%
Here $\mathcal{D}$ is the sampled $(\omega,k)$ grid, $N_{\mathcal{D}}\equiv|\mathcal{D}|$, $\chi_{\mathrm{DATA}}$ is the target, and $\chi_{\mathrm{NODE}}$ is the Neural-ODE prediction. The expression is the grid-averaged Itakura--Saito divergence~\cite{itakuraAnalysisSynthesisTelephony1968}. Its functional form follows Ref.~\cite{Ran:2025vat}, where it was motivated by the Kullback--Leibler divergence between probability distributions generated by field-theoretic partition functions. In the present fermionic problem the positive-argument form is well defined throughout training without introducing an absolute value: $\chi_{\mathrm{DATA}}>0$ by construction, while $\chi_{\mathrm{NODE}}>0$ follows analytically from the spectral positivity of $\mathrm{Im}\,G_{22}$ in Section~\ref{sec:holo-fermions} and of $\mathrm{Im}\,\mathcal{G}_k$ in Appendix~\ref{app:flux}. These results hold for all real network outputs allowed by the parametrizations of $f$, $h$, and $qA_t$. The loss is therefore non-negative and vanishes if and only if $\chi_{\mathrm{NODE}}=\chi_{\mathrm{DATA}}$ at every sampled point.
\subsection{Training Protocol}\label{subsec:hyperparameters}

We employ the Tsitouras 5(4) method~\cite{tsitourasRungeKuttaPairs2011}, an explicit adaptive Runge--Kutta solver well suited to non-stiff problems. We use the relative and absolute ODE tolerances \(\mathrm{rtol}=\mathrm{atol}=10^{-6}\). The radial equations are integrated from the IR cutoff $z_{\mathrm{IR}}=0.999999$, where the in-falling condition is imposed, to the UV cutoff $z_{\mathrm{UV}}=0.0001$, where the boundary spectral function is evaluated. These numerical cutoffs follow the implementation of Ref.~\cite{Ran:2025vat}, but their role is different from the finite HWRG matching scale used there: in the present work $z_{\mathrm{UV}}$ is a fixed regulator for approaching the asymptotic boundary, not a Wilsonian hyperparameter.

Building on the multistage optimization strategy of Ref.~\cite{Ran:2025vat}, we introduce competitive screening and perturbation-based variant generation to concentrate computation on the most promising candidates. In our implementation, under otherwise identical conditions, this modification reduced the wall-clock time of the most expensive examples considerably while giving comparable final losses. The resulting four-stage protocol is as follows.
\begin{description}
\item[Stage 0: Seed Model Selection.] We initialize $10$ candidate models with distinct random seeds and train them in parallel using the Adam optimizer with a learning rate of $10^{-3}$ and a batch size of $64$. Training is halted for all models once the first model completes $500$ epochs, and the $5$ models with the lowest loss are selected as seed models.
\item[Stage 1: Variant Generation and Fine-Tuning.] For each of the $5$ seed models, we retain the original and generate $3$ additional variants by injecting $10\%$ Gaussian noise into the network weights, yielding $20$ models in total. These models then undergo a rapid fine-tuning phase with the Adam optimizer (learning rate $10^{-4}$, batch size $128$, $100$ epochs), after which the top $10$ models are retained.
\item[Stage 2: Deep Optimization.] The $10$ surviving models are trained further to settle into deeper loss minima, using the Adam optimizer with a learning rate of $10^{-5}$ and a batch size of $128$ or $256$, depending on the experiment, for up to $400$ epochs. The top $5$ models are selected for the final stage. The robustness reruns newly reported in Appendix~\ref{app:robustness} use a batch size of $128$.
\item[Stage 3: Precision Polishing (BFGS).] The remaining $5$ models undergo second-order optimization via the BFGS algorithm to achieve high numerical precision. The optimization proceeds for up to $2000$ steps, with both the relative and absolute optimizer stopping tolerances set to $10^{-9}$.
\end{description}

Unless otherwise stated, the tabulated loss and mean relative error (MRE) below are evaluated using the model with the lowest loss among the five candidates from the final training stage. For a quantity $X$, we define the MRE as
\begin{equation}
\mathrm{MRE}_{X}
=\frac{1}{N_X}\sum_{i\in\mathcal{I}_X}
\left|
\frac{X_i^{\mathrm{pred}}-X_i^{\mathrm{ref}}}
{X_i^{\mathrm{ref}}}
\right|,
\label{eq:mre}
\end{equation}
where $\mathcal{I}_X$ is the corresponding numerical evaluation grid restricted to points with $X_i^{\mathrm{ref}}\neq0$, and $N_X\equiv|\mathcal{I}_X|$; the values below are reported as $100\,\mathrm{MRE}_X$ in percent. For spectral comparisons, $X^{\mathrm{pred}}=\chi_{\mathrm{NODE}}$, $X^{\mathrm{ref}}=\chi_{\mathrm{DATA}}>0$, and $\mathcal{I}_X=\mathcal{D}$. The benchmark bulk-field MREs are evaluated on $1000$ uniformly spaced points over $z\in[z_{\mathrm{UV}},z_{\mathrm{IR}}]$.

For the bulk-field figures, each of the five final candidates is mapped independently, first to the constant-$h$ coordinate $u$ and then, when a common-temperature comparison is required, to the coordinate $v$ defined below. In the PLL figures, $u$ is only an intermediate coordinate: the panels show the original $z$-frame candidates and the final $v$-frame profiles, but not the $u$-frame curves. Each mapped candidate is evaluated on the same $1000$-point $v$ grid, after which the ensemble mean and pointwise sample standard deviation are calculated at each $v$ point. The controlled synthetic-reconstruction figure additionally shows the $u$-frame profiles and applies the same procedure separately on the displayed $u$ and $v$ grids. These ensemble statistics are used only for visualization. Consistent with the convention above, every reported bulk-field MRE is evaluated for the lowest-loss candidate: on the original $z$ grid for the benchmarks, and after the corresponding coordinate maps on the common $1000$-point grid for the PLL and controlled synthetic reconstructions. The sample variance uses the denominator $N-1$ for $N=5$. The shaded bands quantify dispersion among the five retained optimization candidates; they are not confidence intervals for the reconstructed geometry. Thin gray curves denote the original $z$-frame candidates, and the lowest-loss candidate in the displayed mapped frame is highlighted separately.

\section{Numerical Results}\label{sec:examples}
Our goal in this section is to test the proposed fermionic holographic machine-learning framework. We use the following three examples as controlled data generators. The Einstein--Maxwell and Gubser--Rocha models provide synthetic holographic spectra with known bulk solutions, allowing us to quantify the reconstruction accuracy against the exact metric functions and gauge potential. The extended PLL case instead uses an analytic phenomenological target with parameters calibrated by cuprate ARPES measurements. Its purpose is not to infer bulk fields directly from measured MDC intensities or to verify a known bulk solution, but to implement the geometric test posed in the Introduction by searching for effective representatives of the extended-PLL interaction sector within the prescribed ansatz.

In Sections~\ref{subsec:em} and~\ref{subsec:gr} we use the first ansatz, Eqs.~\eqref{ads4}--\eqref{eq:qA_ads4}. The second ansatz, Eq.~\eqref{cadsr2}, is reserved for Section~\ref{subsec:pll-k}, where it is applied to the extended PLL data.
\subsection{Einstein--Maxwell Theory}\label{subsec:em}

In the pioneering work on holographic non-Fermi liquids~\cite{liuNonFermiLiquidsHolography2011}, the bulk theory is governed by the Einstein--Maxwell action%
\begin{equation}
S=\int d^{4}x\sqrt{-g}\left[ \mathcal{R}+6-F_{\mu \nu }F^{\mu \nu }\right] ,
\end{equation}%
where we have set the AdS radius and the effective gauge coupling to unity. The equations of motion admit a charged black hole solution in asymptotically AdS$_4$ spacetime, with metric functions and gauge potential%
\begin{equation}
f(z)=1-z^{3}+Q^{2}\left( z^{4}-z^{3}\right) ,\;h(z)=1,\;A_{t}(z)=Q(1-z).
\label{fhAEM}
\end{equation}%
Here $Q$ ranges over $[0,\sqrt{3}]$.

\begin{figure}[tbph]
\centering
\includegraphics[width=1\textwidth]{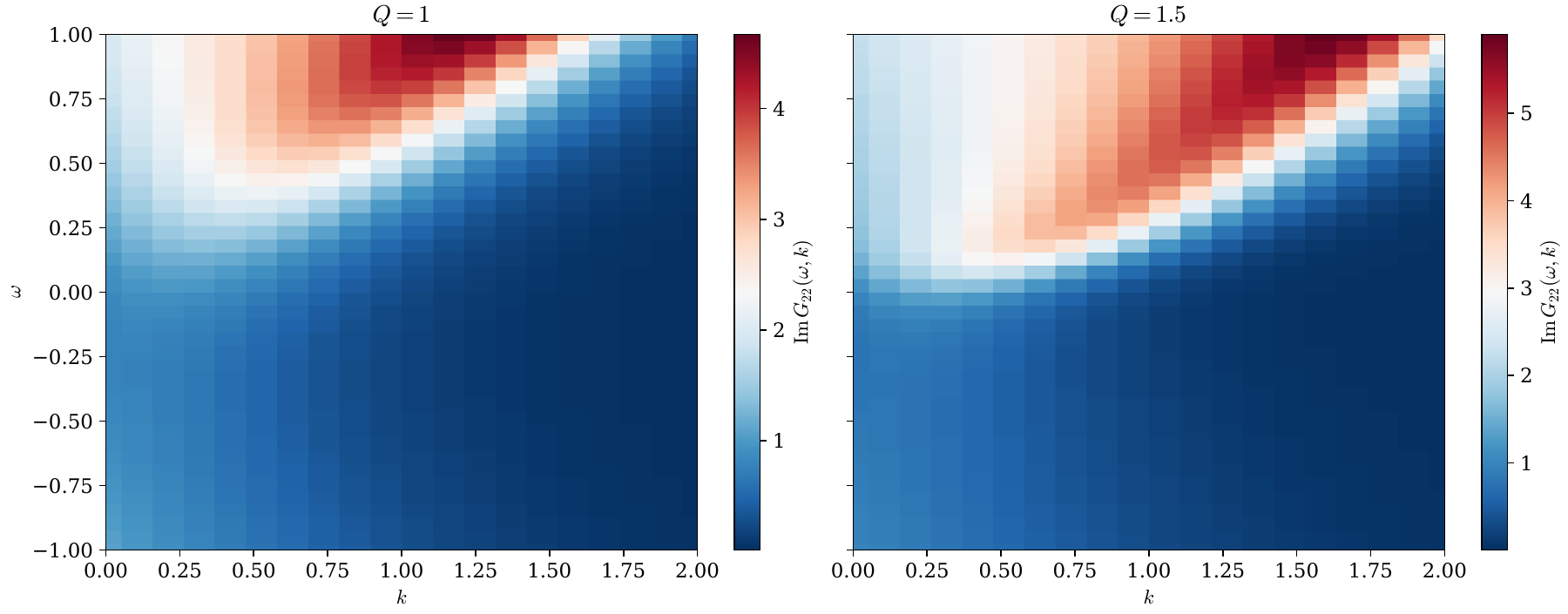}
\caption{Heatmap of the spectral function in the Einstein--Maxwell theory, for $Q=1$ (left) and $Q=1.5$ (right).}
\label{fig:heatmap}
\end{figure}

\begin{figure}[tbph]
\centering
\includegraphics[width=1.0\textwidth]{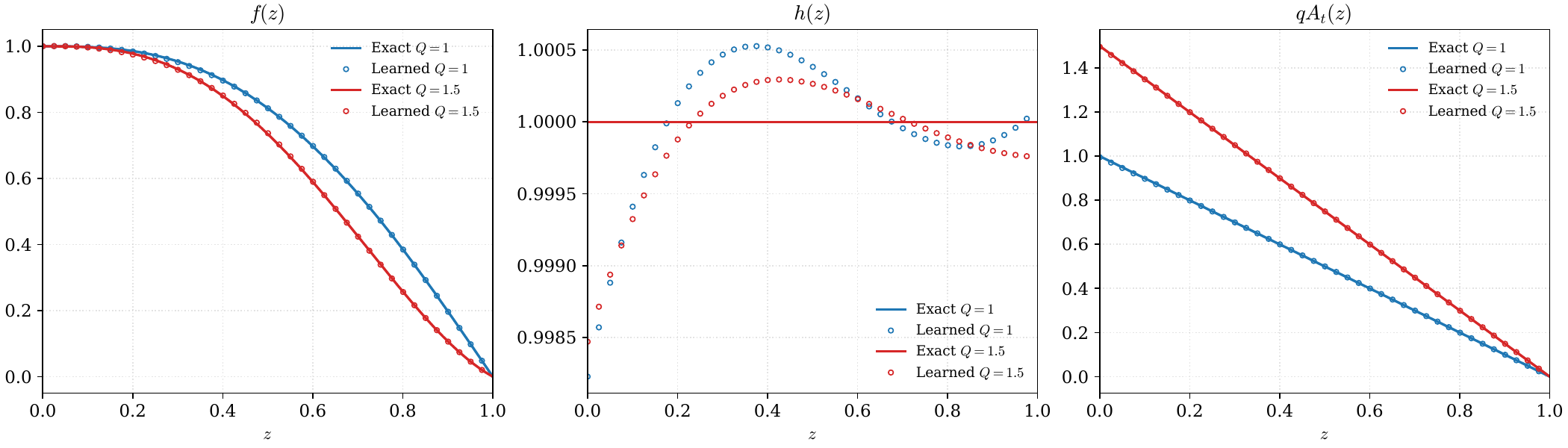}
\caption{Reconstruction in the Einstein--Maxwell theory: comparison
between the learned bulk fields $f(z),h(z)$ and $qA_{t}(z)$ (markers) and the analytical RN solutions (solid lines), for $Q=1$ and $Q=1.5$.}
\label{fig:rn_results}
\end{figure}

To generate the training data we set $q=1$ and the fermion mass $m=1/4$ as representative values.\footnote{Other values of $q$ and $m$ are also admissible; see the other two examples in this section.} %
We take $Q=1$ and $Q=1.5$ as two benchmark cases. Substituting Eq.~(\ref{fhAEM}) into the holographic flow equation~(\ref{Hflow}) with the in-falling boundary condition $\xi_{-}(z_{\mathrm{IR}})=i$, we integrate the flow equation to the UV boundary and obtain the spectral function $\mathrm{Im}\,G_{22}=\mathrm{Im}\,z_{\mathrm{UV}}^{-2m}\xi_{-}(z_{\mathrm{UV}})$. By uniformly sampling 40 points in $\omega \in [-1,1]$ and 20 points in $k\in [0,2]$, we generate a total of 800 data points; the corresponding heatmap is shown in Fig.~\ref{fig:heatmap}. 
The training results are summarized in Fig.~\ref{fig:rn_results} and Table~\ref{tab:rn_gr_results}: the learned metric functions and $qA_t$ match the exact analytical solutions with high precision.

\subsection{Gubser--Rocha Model}\label{subsec:gr}

\begin{figure}[htbp]
\centering
\includegraphics[width=1\textwidth]{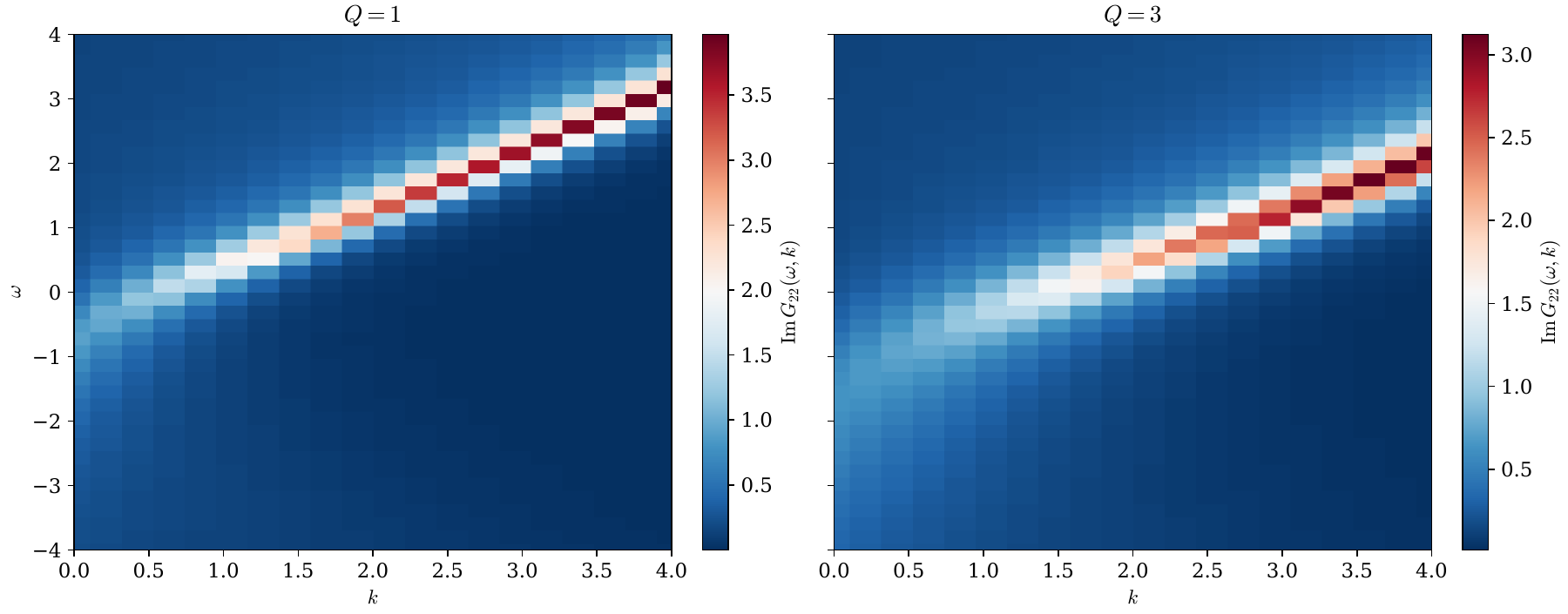}
\caption{Heatmap of the spectral function in the Gubser--Rocha model, for $Q=1$ (left) and $Q=3$ (right).}
\label{fig:heatmapGR}
\end{figure}

\begin{figure}[htbp]
\centering
\includegraphics[width=1\textwidth]{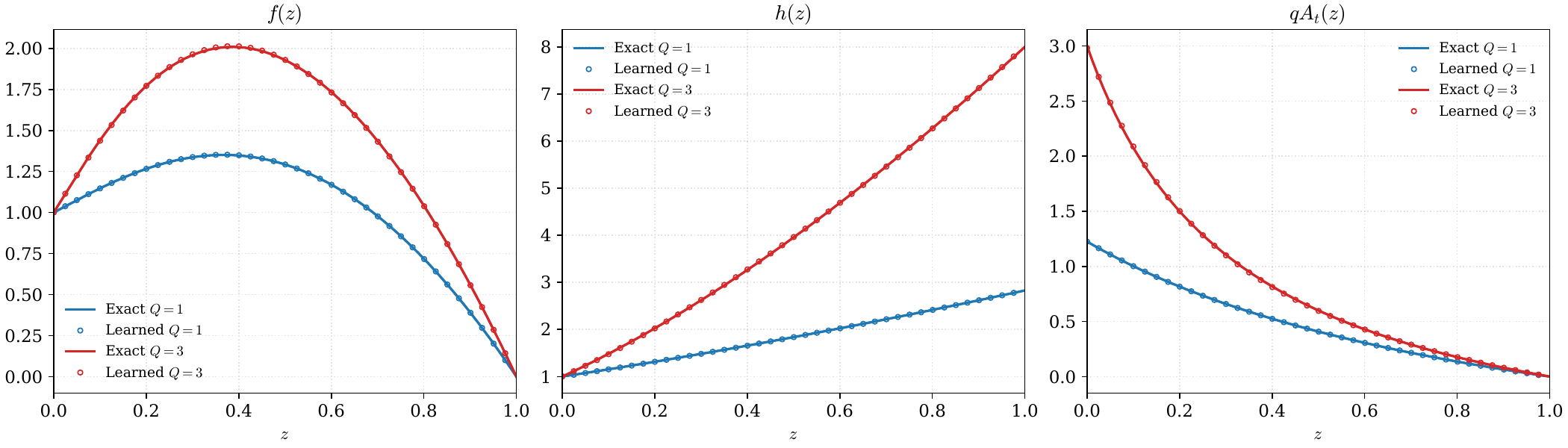}
\caption{Reconstruction in the Gubser--Rocha model: comparison between the learned bulk fields $f(z),h(z)$ and $qA_{t}(z)$ (markers) and the analytical solutions (solid lines), for $Q=1$ and $Q=3$.}
\label{fig:fhaGR}
\end{figure}

The Gubser--Rocha model is a widely studied holographic model for strange metals~\cite{gubserPeculiarPropertiesCharged2010,zaanenHolographicDualityCondensed2015}, in which a dilaton $\phi$ is coupled to the Maxwell field in a specific way,%
\begin{equation}
S=\int d^{4}x\sqrt{-g}\left[ \mathcal{R}-\frac{\partial _{\mu }\phi \partial ^{\mu
}\phi }{2}+6\cosh \left( \frac{\phi }{\sqrt{3}}\right) -\frac{e^{\phi /\sqrt{%
3}}}{4}F_{\mu \nu }F^{\mu \nu }\right] .
\end{equation}%
The equations of motion admit a charged black hole solution in AdS$_4$, whose metric functions and gauge potential are
\begin{align}
f(z) &=(1-z)\frac{1+z+3Qz+\left( 1+3Q+3Q^{2}\right) z^{2}}{\left(
1+Qz\right) ^{3/2}}, \\
h(z) &=\left( 1+Qz\right) ^{3/2},\;A_{t}(z)=(1-z)\frac{\sqrt{3Q(1+Q)}}{1+Qz}%
,
\end{align}%
where $Q$ is a positive parameter. We generate the training data with $m=-1/4$ and $q=1/2$, and take $Q=1$ and $Q=3$ as two benchmark cases. As in Section~\ref{subsec:em}, $40 \times 20$ data points are sampled in $\omega \in [-4,4]$ and $k\in [0,4]$, see Fig.~\ref{fig:heatmapGR}. The training results are shown in Fig.~\ref{fig:fhaGR} and Table~\ref{tab:rn_gr_results}, confirming the high reconstruction accuracy.
\begin{table}[htbp]
\caption{The performance of Neural ODEs on the Einstein--Maxwell (EM) and
Gubser--Rocha (GR) models. The columns report the training loss and the MRE of each bulk field $f(z),h(z),qA_t(z)$ against its analytical value.}
\label{tab:rn_gr_results}\centering
\begin{tabular}{lccccc}
\toprule
\textbf{Model} & \textbf{$Q$} & \textbf{Loss}
& \textbf{$f(z)$ MRE} & \textbf{$h(z)$ MRE}
& \textbf{$qA_t(z)$ MRE} \\
\midrule \textbf{EM} & 1 & $2\times 10^{-10}$ & $0.1\%$ & $0.03\%$ & $0.1\%$
\\ 
\textbf{EM} & 1.5 & $1\times 10^{-10}$ & $0.2\%$ & $0.03\%$ & $0.04\%$ \\ 
\textbf{GR} & 1 & $1\times 10^{-9}$ & $0.04\%$ & $0.01\%$ & $0.08\%$ \\
\textbf{GR} & 3 & $2\times 10^{-10}$ & $0.1\%$ & $0.2\%$ & $0.3\%$ \\
\bottomrule
\end{tabular}%
\end{table}

\subsection{Application to Extended-PLL model}\label{subsec:pll-k}

The extended PLL scaling is inspired by the Gubser--Rocha IR Green's function in the semi-holographic framework. As shown in Section~\ref{subsec:grav-contribution} and Appendix~\ref{app:ir-green}, the associated Dirac equation reduces to that of a geometry conformal to the AdS$_2\times\mathbb{R}^2$ black hole. The fixed analytical solution grounds the momentum-dependent exponent $\alpha(k)$ but underestimates the thermal contribution to the self-energy at elevated temperatures~\cite{mauriGaugegravityDualityComes2024}. To test the proposed geometric explanation, we adopt the broader class in Eq.~\eqref{cadsr2}, with trainable metric functions and $qA_t$, and use the extended PLL model as the target, which provides an accurate phenomenological description of the observed doping and temperature dependence~\cite{smitMomentumdependentScalingExponents2024,mauriGaugegravityDualityComes2024}. The inverse problem then asks whether, and over what ranges of doping and temperature, this ARPES-calibrated interaction self-energy admits an effective realization beyond the fixed Gubser--Rocha IR benchmark.

Translating the extended-PLL input into suitable training data requires addressing three points. First, as emphasized in Refs.~\cite{smitMomentumdependentScalingExponents2024,mauriGaugegravityDualityComes2024} and discussed in Section~\ref{subsec:grav-contribution}, the bulk holographic fermion is dual to a composite fermionic operator describing hole excitations. Since ARPES measures the electronic response, a particle-hole conjugation must be performed on this composite operator,%
\begin{equation}
\mathcal{O}_{h}(\omega ,k)\rightarrow \mathcal{O}_{e}^{\dagger }(-\omega
,2k_{F}-k).
\end{equation}%
Second, under this conjugation the semi-holographic self-energy~(\ref{sigma}) transforms as~\cite{mauriGaugegravityDualityComes2024}
\begin{equation}
\begin{aligned}
\Sigma _{h}(\omega ,k)
&=-g_{k}^{2}\mathcal{G}_{k}(\omega )
\\
&\rightarrow g_{2k_{F}-k}^{2}\mathcal{G}_{2k_{F}-k}^{\ast }(-\omega )
=\Sigma _{e}(\omega ,k).
\end{aligned}
\end{equation}%
Consistent with the MDC reduction in Eq.~\eqref{eq:semi-holo-mdc}, one can identify the positive extended-PLL linewidth contribution with $\Sigma_{\mathrm{E}}^{\prime \prime} \equiv -\mathrm{Im}\,\Sigma_{e}$, giving%
\begin{equation}
\Sigma _{\mathrm{E}}^{\prime \prime }(\omega ,k)=g_{2k_{F}-k}^{2}\mathrm{Im}\,\mathcal{G}_{2k_{F}-k}(-\omega ).
\label{sigmaH}
\end{equation}%
The momentum-dependent coupling $g_{2k_F-k}^{2}$ is fixed phenomenologically to match the measured peak width near the Fermi surface and is not determined by holography. Third, the correspondence between the bulk variables $\omega$ and $k$ and the dimensionful experimental observables must be specified.

With these points in mind, we define our training data as:%
\begin{equation}
\begin{aligned}
\chi _{\mathrm{DATA}}(\omega ,k)
&=\frac{\Sigma _{\mathrm{E}}^{\prime \prime }(\omega ,k)}{%
\Sigma _{\mathrm{E}}^{\prime \prime }(\omega _{0},k)}
&=\frac{[(\frac{\hbar \omega }{k_{B}T}%
)^{2}+\beta ^{2}]^{\alpha \left( k\right) }}{[(\frac{\hbar \omega _{0}}{%
k_{B}T})^{2}+\beta ^{2}]^{\alpha \left( k\right) }},
\end{aligned}
\label{chidata}
\end{equation}%
where $\omega_0$ is the reference frequency and $\Sigma _{\mathrm{E}}^{\prime \prime }$ comes from Eqs.~\eqref{sigmaHpll} and~\eqref{ak}. Hereafter, $\chi_{\mathrm{DATA}}$ denotes the normalized extended-PLL target in Eq.~\eqref{chidata}, whose parameters ($\alpha,\beta$) are fixed by the ARPES analysis of Ref.~\cite{smitMomentumdependentScalingExponents2024}. Following this reference, we consider ARPES-derived parameter sets for single-crystal Bi2201 [$\mathrm{(Pb,Bi)}_2\mathrm{Sr}_{2-x}\allowbreak\mathrm{La}_x\allowbreak\mathrm{CuO}_{6+\delta}$] samples spanning the underdoped and overdoped regimes, using the sample labels adopted therein (see Table~\ref{tab:errors} for the complete list). On the Neural ODE side, the output takes the form%
\begin{equation}
\chi _{\mathrm{NODE}}(\omega ,k)=\frac{\mathrm{Im}\,\mathcal{G}_{2k_{F}-k}(-\omega )}{%
\mathrm{Im}\,\mathcal{G}_{2k_{F}-k}(-\omega _{0})}=\frac{\left. \dfrac{z^{-2\nu _{k}}\,\mathrm{Im}\left( y_{+}\bar{z}_{-}\right) }{\left| y_{+}+z_{-}\right| ^{2}}\right\vert _{\substack{ \omega \rightarrow -\omega  \\ %
k\rightarrow 2k_{F}-k}}}{\left. \dfrac{z^{-2\nu _{k}}\,\mathrm{Im}\left( y_{+}\bar{z}_{-}\right) }{\left| y_{+}+z_{-}\right| ^{2}}%
\right\vert _{\substack{ \omega \rightarrow -\omega _{0}  \\ k\rightarrow
2k_{F}-k}}}.
\label{chinode}
\end{equation}%
Through these two expressions, the particle-hole conjugation is properly incorporated. The normalization cancels $\lambda$, $\omega_N$, and the undetermined coupling $g_{2k_F-k}^{2}$, while the additive contribution $G_0$ is absent because the target contains only $\Sigma_{\mathrm E}^{\prime\prime}$. Absolute intensity, measurement uncertainties, and deviations from the extended-PLL form are therefore not included. The normalized construction thus isolates the geometric direction; temperature dependence in $g_k^2$ does not affect this spectral-shape test, and additional contributions to $G_0$ are not tested. 

It remains to specify the unit map. In the dimensionless bulk units used here, the scaled Hawking temperature is defined as
\begin{equation}
T_{H}=-\frac{f^{\prime }\left( 1\right) }{4\pi }.  \label{TH}
\end{equation}%
The bulk frequency $\omega$ is related to the dimensionless boundary ratio $\hbar\omega/(k_B T)$ via~\cite{Ran:2025vat}
\begin{equation}
\omega \rightarrow \frac{-f^{\prime }\left( 1\right) }{4\pi }\frac{\hbar
\omega }{k_{B}T},
\end{equation}%
with $\hbar\omega$ and $k_B T$ expressed in eV. We quote the learned $qA_t$ in the same energy units. The unit of $k$ may be chosen freely thanks to the scaling symmetry~(\ref{ss2}); we adopt \AA$^{-1}$ for convenience. Appendix~\ref{app:units} gives the SI unit restoration and the mapping from bulk variables to experimental observables.

We set the reference frequency to $\hbar\omega_0=-0.01\text{ eV}$.\footnote{If $\chi_{\mathrm{NODE}}=\chi_{\mathrm{DATA}}$ throughout the fitting window for one choice of $\omega_0$, the equality is preserved after normalizing both spectra at any other reference frequency within that window. For the finite-accuracy fits reported here, any residual dependence is controlled by the fitting error. We choose the endpoint value for convenience.} For each sample at a given temperature, we generate 800 uniformly distributed data points. The sampled windows are $\hbar\omega \in [-0.3,-0.01]\,\mathrm{eV}$ and $k \in [0,0.7]\,\text{\AA}^{-1}$. We restrict the analysis to the negative-frequency domain accessible to standard ARPES and to the positive node branch by symmetry; see Fig.~2 of Ref.~\cite{smitMomentumdependentScalingExponents2024}.

The training results are organized as follows.

\begin{figure}[htbp]
\centering
\includegraphics[width=1\textwidth]{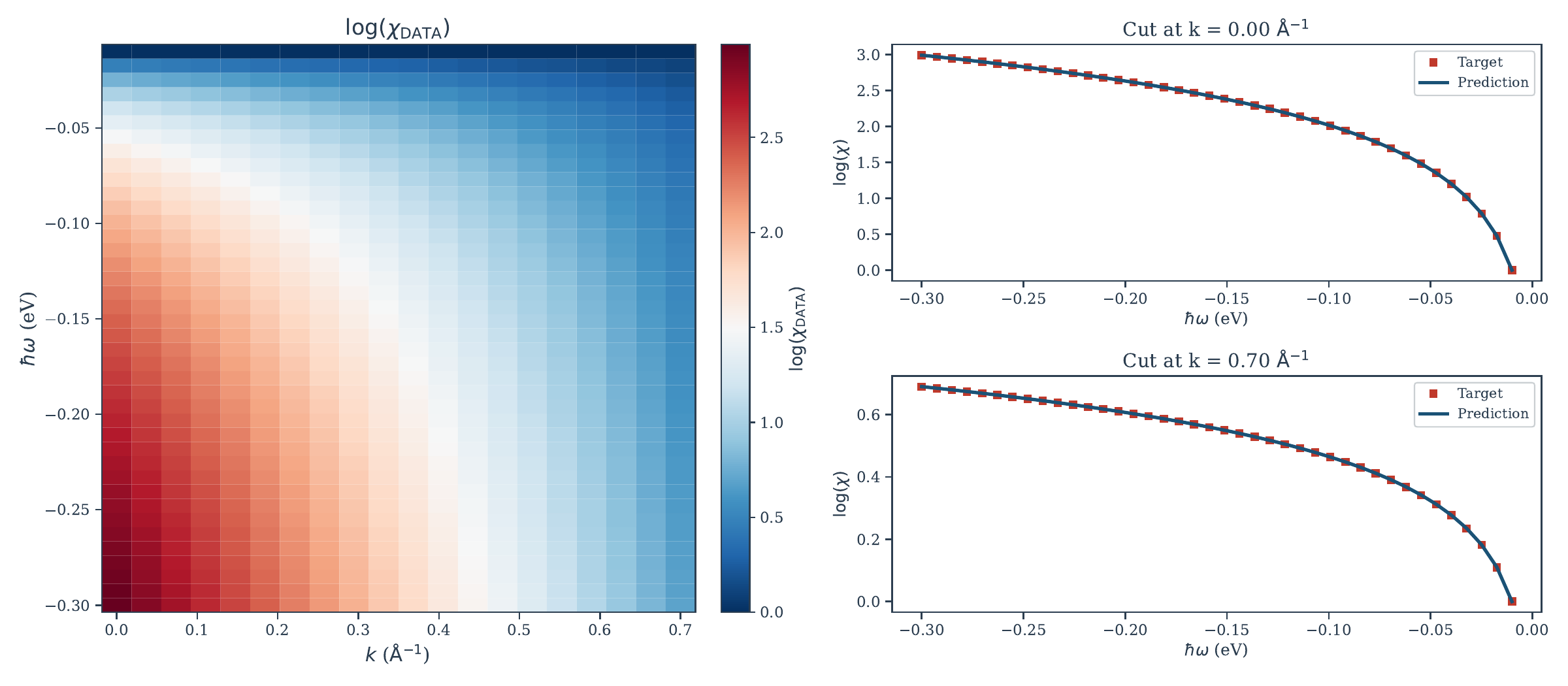}
\caption{Normalized extended-PLL target and Neural-ODE prediction for the near-optimally doped UD32K sample at $T=8\,\mathrm{K}$. Left: the target $\protect\chi_{\mathrm{DATA}}(\protect\omega,k)$ generated from Eq.~(\protect\ref{chidata}). Right: the comparison between the target and predicted spectra at $k=0$ (top) and $k=0.7\,\text{\AA}^{-1}$ (bottom).}
\label{fig:gwk_pll}
\end{figure}

1. \textbf{Near-optimal doping at low temperature:} We first focus on the near-optimally doped UD32K sample at $T=8\,\mathrm{K}$. The PLL parameters ($\alpha = 0.51$, $\beta = 3.39$) and the Fermi momentum ($k_F = 0.455\text{ \AA}^{-1}$) are taken from Fig.~1 and Supplementary Fig.~3 of Ref.~\cite{smitMomentumdependentScalingExponents2024}, respectively. The training dataset and the Neural ODE prediction are shown in Fig.~\ref{fig:gwk_pll}.
The Neural ODE prediction closely matches the training data, achieving a loss of $2\times10^{-8}$ (Table~\ref{tab:errors}).

\begin{figure}[tbph]
\centering
\includegraphics[width=1\textwidth]{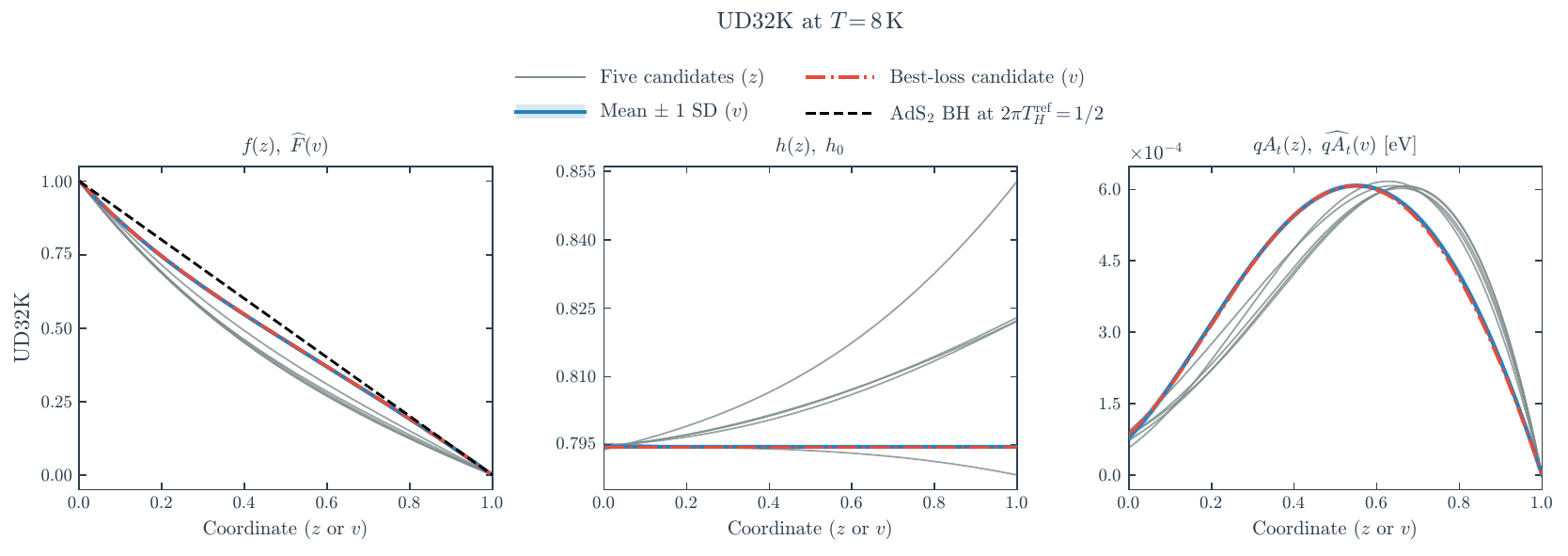}
\caption{Learned effective bulk fields for the near-optimally doped UD32K sample at $T=8\,\mathrm{K}$. The panels show the blackening factor, spatial metric, and $qA_t$ profile. Thin gray curves show the five final candidates in the original $z$ coordinate. Blue curves and very narrow bands show the mean and pointwise 1-SD spread after each candidate is mapped through $z\to u\to v$ to the common representative $2\pi T_H^{\mathrm{ref}}=1/2$; red dash-dotted curves show the lowest-loss candidate in that frame. The black dashed curve is the exact $\mathrm{AdS}_2$ black hole at $T_H^{\mathrm{ref}}$. The mapped blackening factors share its endpoint values and horizon slope by construction but show a systematic interior deviation.}
\label{fig:fha_pll}
\end{figure}

The bulk functions learned by the fully trainable model are shown in Fig.~\ref{fig:fha_pll}. Before interpreting them, two analytic redundancies must be accounted for. First, the coordinate/Weyl redundancy of the pair $\{f,h\}$ is removed by mapping each candidate separately to the constant-$h$ coordinate $u$, as described in Appendix~\ref{app:F}. Second, even in this frame, the normalized spectra do not determine the scaled Hawking temperature $T_H$. Appendix~\ref{app:th_degeneracy} shows that any regular constant-$h$ blackening factor, together with the appropriately transported $qA_t$, belongs to a temperature-map orbit with the same normalized spectrum. We therefore select the common representative $2\pi T_H^{\mathrm{ref}}=1/2$ as a frame convention, rather than as an independently inferred temperature.

\begin{figure}[htbp]
\centering
\includegraphics[width=1\textwidth]{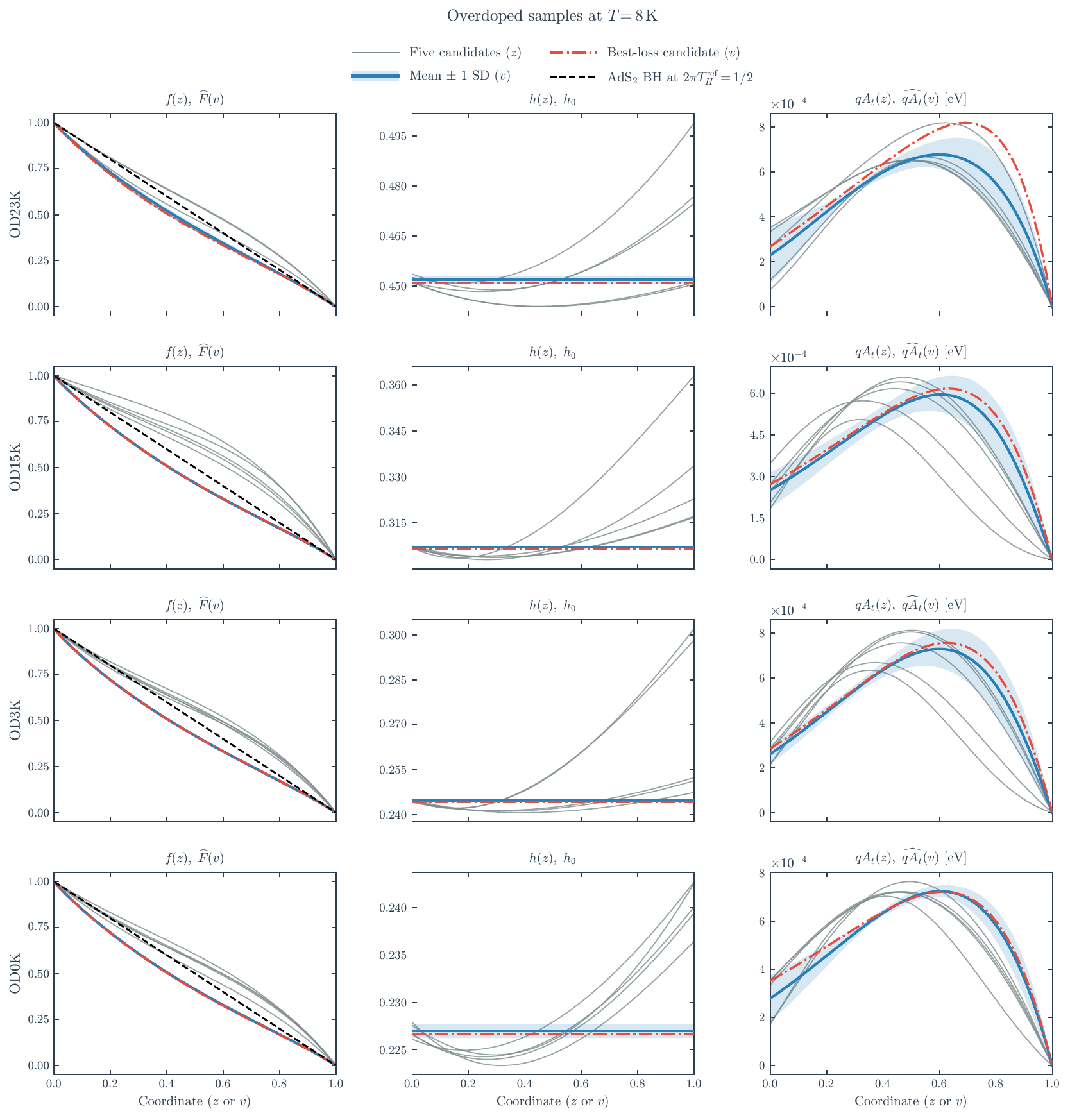}
\caption{Effective bulk fields for the overdoped samples (OD23K, OD15K, OD3K, OD0K) at $T=8$~K. Each row corresponds to one sample, and the columns show the blackening factor, spatial metric, and gauge potential. The notation and two-step mapping follow Fig.~\ref{fig:fha_pll}. The black dashed curves are the exact $\mathrm{AdS}_2$ reference at the common $2\pi T_H^{\mathrm{ref}}=1/2$. The mapped blackening factors share its endpoint values and horizon slope by construction but exhibit sample-dependent interior deviations, while the gauge potential remains negligible ($qA_t<10^{-3}\,\mathrm{eV}$).}
\label{fig:fha_OD}
\end{figure}

We compare with the exact $\mathrm{AdS}_{2}$ black-hole form in this frame,
\begin{equation}
F_{\mathrm{AdS}_2}(v)=(1-v)\bigl[1+v(4\pi T_H-1)\bigr],
\label{eq:F_ads2bh}
\end{equation}
which is the unique solution of $\mathcal{R}_2=-2$ subject to $F_{\mathrm{AdS}_2}(1)=0$ and $T_H=-F_{\mathrm{AdS}_2}'(1)/(4\pi)$. For candidate $i$, we use its scaled Hawking temperature in the $u$-frame, $T_{H,i}=-F_i'(1)/(4\pi)$, and map the candidate to the common $v$-frame at $T_H^{\mathrm{ref}}$ with $\eta_i=T_H^{\mathrm{ref}}/T_{H,i}$ using Eqs.~(\ref{eq:diffeo}) and~(\ref{eq:At_transport_temp}) of Appendix~\ref{app:th_degeneracy}. In Figs.~\ref{fig:fha_pll}, \ref{fig:fha_OD}, and~\ref{fig:fha_highT}, the means and pointwise 1-SD bands are formed only after this candidate-by-candidate map, and the dashed curve is the exact Eq.~(\ref{eq:F_ads2bh}) at $T_H^{\mathrm{ref}}$.

The three learned functions can now be interpreted consistently. First, at the selected $2\pi T_H^{\mathrm{ref}}=1/2$, every mapped blackening factor satisfies $\widehat F(0)=1$, $\widehat F(1)=0$, and $\widehat F'(1)=-1$, just as the exact reference $F_{\mathrm{AdS}_2}(v)=1-v$; these endpoint and horizon-slope agreements follow from the boundary conditions and the common-temperature frame rather than from an independent fit. The mapped $\widehat F(v)$ profile of the lowest-loss candidate has a broadly similar monotonic shape and an MRE of approximately $6\%$ relative to $1-v$, but it lies systematically below the reference over much of the interior radial interval. Second, the learned spatial factor gives $\nu_{k_F}=k_F/\sqrt{h_0}\approx0.510$, consistent with the input PLL exponent $\alpha=0.51$. Third, for the lowest-loss fully trainable candidate, the charge-weighted gauge potential is very small in magnitude, $qA_t\sim10^{-4}\,\mathrm{eV}$, with $\left|qA_t(z)/(\hbar\omega)\right|$ typically of order $10^{-3}$--$10^{-2}$ over the data window.

These results should be interpreted with the limited low-temperature sensitivity in mind. At $T=8\,\mathrm{K}$, the thermal scale $\beta k_BT\approx2.34\,\mathrm{meV}$ is smaller than the lowest sampled $|\hbar\omega|=10\,\mathrm{meV}$, so most of the target is dominated by its power-law behavior. The agreement $\nu_{k_F}\approx\alpha$ is therefore strongly constrained, but should be viewed as a consistency check of the massless effective model chosen on the basis of the target scaling and boundary regularity rather than as an independent determination of the probe mass. A pure power law fixes this boundary exponent but does not by itself select an $\mathrm{AdS}_2$ blackening factor. The lower-energy points in the sampled window carry larger finite-temperature corrections and help constrain the combined spectral response of the metric and gauge profiles beyond the leading power law. On the same grid, the pure-power contribution $\chi_{\mathrm{power}}=|\omega/\omega_0|^{2\alpha(k)}$ gives a loss of approximately $3\times10^{-4}$, the fixed analytical IR Gubser--Rocha benchmark gives $3\times10^{-5}$, and the fully trainable model gives $2\times10^{-8}$. The analytical benchmark is thus a reasonable low-temperature approximation, while the fully trainable model, on the sampled training grid, more accurately reproduces the smaller thermal-crossover features of the analytic extended-PLL target. 

Restricted fits show that most of the improvement over the fixed analytical benchmark can be obtained by allowing either the geometry or the gauge potential to vary separately. The fully trainable model nevertheless achieves a further reduction in loss. Independently, because a nonzero $qA_t$ generically introduces a frequency asymmetry, the known even-frequency structure of the PLL target favors $qA_t\simeq0$ within the adopted ansatz.

2. \textbf{Conformal factor and thermodynamics:} The accurate spectral fit and the smooth learned profiles within the adopted parametrization show that the present framework admits smooth representatives of the effective conformal-class Dirac background compatible with the normalized extended-PLL spectra for UD32K at low temperatures. With the chosen massless effective probe, the spectral function is independent of the conformal factor. Boundary regularity alone does not determine this factor or its temperature dependence. Thermodynamic quantities that depend on this factor are therefore not fixed by the fermionic spectra. For instance, in an Einstein-gravity realization the Bekenstein--Hawking entropy density for the ansatz~\eqref{cadsr2} is given by
\begin{equation}
s(T)\propto \sqrt{g_{xx}g_{yy}}\bigg|_{z=1}=\Omega(1,T)\,h(1,T).
\label{eq:entropy}
\end{equation}
In the constant-$h$ gauge of Appendix~\ref{app:F}, the spectra constrain $h(1,T)$ through $\nu_k$, but the conformal factor $\Omega(1,T)$ is entirely invisible to the massless probe. Consequently, determining this entropy density and its contribution to the electronic specific heat requires independent macroscopic input.

3. \textbf{Doping dependence:} We next examine the four overdoped samples at $T=8\,\mathrm{K}$. As shown in Table~\ref{tab:errors}, the framework accurately reproduces the normalized extended-PLL targets on the sampled training grids for all five samples. The loss is smallest for UD32K, the sample closest to the marginal Fermi liquid point ($\alpha=1/2$), and increases monotonically toward the overdoped side. Even for OD0K, however, the loss is only $\approx4\times10^{-7}$ and the $\chi$ MRE is $\approx0.08\%$; the deterioration in spectral accuracy is therefore mild over the studied range. For every sample, the learned constant spatial factor also reproduces the input PLL exponent through $\nu_{k_F}=k_F/\sqrt{h_0}\approx\alpha$.

Turning to the radial profiles, $qA_t$ remains below $10^{-3}\,\mathrm{eV}$ in the four fully trainable overdoped fits shown in Fig.~\ref{fig:fha_OD}. After applying the same candidate-wise mapping as for UD32K, the MREs of the lowest-loss $\widehat F(v)$ profiles relative to the exact $\mathrm{AdS}_2$ reference are approximately $11\%$--$13\%$, larger than the UD32K value reported above.

\begin{figure}[htbp]
\centering
\includegraphics[width=1\textwidth]{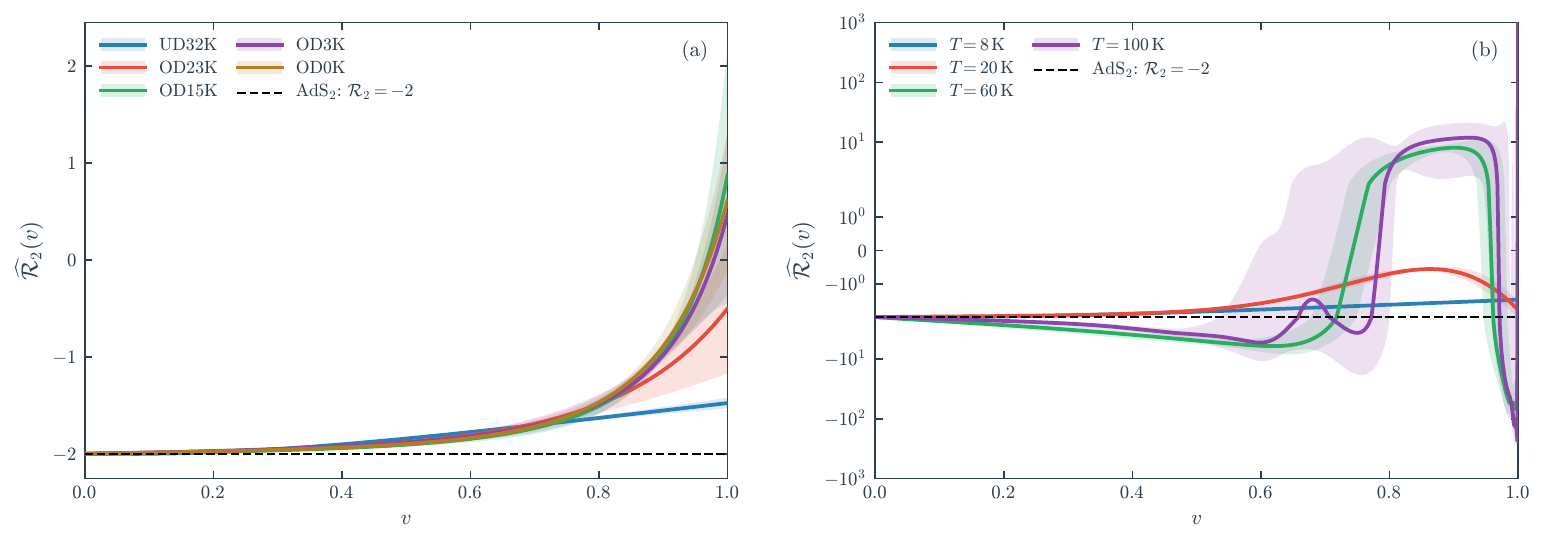}
\caption{Scalar curvature $\widehat{\mathcal{R}}_2(v)$ after mapping each candidate to the common representative $2\pi T_H^{\mathrm{ref}}=1/2$. (a) Five samples at $T=8\,\mathrm{K}$. (b) UD32K at $T=8$, $20$, $60$, and $100\,\mathrm{K}$. Solid curves and shaded bands denote the five-candidate mean and pointwise 1-SD spread; the dashed line denotes $\mathcal{R}_2=-2$.}
\label{fig:curvature_lowT}
\end{figure}

\begin{figure}[htbp]
\centering
\includegraphics[width=1\textwidth]{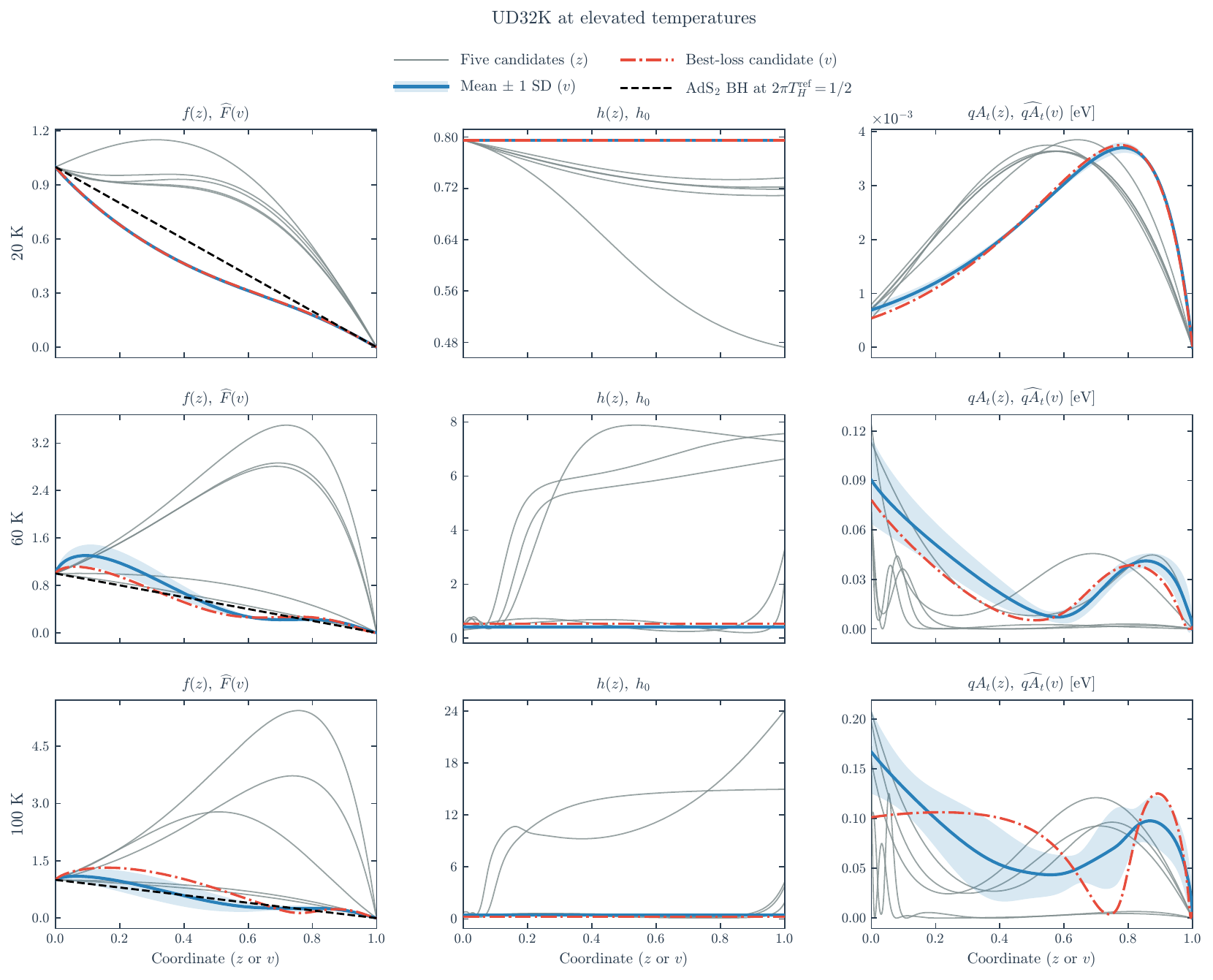}
\caption{Bulk profiles obtained within the prescribed ansatz for the UD32K sample at elevated temperatures ($T=20, 60, 100$~K), with the same notation as Fig.~\ref{fig:fha_pll}. All candidates are mapped to the common $2\pi T_H^{\mathrm{ref}}=1/2$ before calculating the mean and 1-SD bands. At $60$ and $100$~K, $\widehat F(v)$ shows increasing residual deviation and dispersion relative to the exact $\mathrm{AdS}_2$ reference (dashed curve); at $100$~K, the mean $\widehat{qA}_t(v)$ reaches $\sim0.1\,\mathrm{eV}$.}
\label{fig:fha_highT}
\end{figure}

As a local diagnostic of the gauge-fixed two-dimensional $(t,v)$ representative, we evaluate its scalar curvature after both transformations,
\begin{equation}
\widehat{\mathcal{R}}_2(v)=-v^2\widehat F''(v)+2v\widehat F'(v)-2\widehat F(v).
\label{eq:R2_curvature}
\end{equation}
Operationally, we evaluated the curvatures in the constant-$h$ frame using analytic derivatives of the learned profiles and then transported to the common $v$-frame. The UV value $\widehat{\mathcal{R}}_2(0)=-2$ follows automatically from $\widehat F(0)=1$ and is therefore not independent evidence for an $\mathrm{AdS}_2$ geometry. Figure~\ref{fig:curvature_lowT}(a) shows that the curvature can remain close to $-2$ away from the horizon, but departs from it toward the horizon, with order-one deviations for several samples. Thus the profile-level resemblance of $\widehat F(v)$ to the $\mathrm{AdS}_2$ reference does not establish an exact or approximately near-horizon $\mathrm{AdS}_2$ geometry.

4. \textbf{High-temperature test:} To assess the temperature dependence, we repeat the UD32K analysis at $T=20$, $60$, and $100\,\mathrm{K}$ using the same ARPES-calibrated extended-PLL target. As the temperature rises, the thermal crossover enters the sampled energy window; at $100\,\mathrm{K}$, $\beta k_BT\approx29\,\mathrm{meV}$ exceeds the lowest sampled $|\hbar\omega|=10\,\mathrm{meV}$. Correspondingly, the best negative-frequency loss increases by roughly three orders of magnitude, from $2\times10^{-8}$ at $8\,\mathrm{K}$ to $3\times10^{-5}$ at $100\,\mathrm{K}$, while the spectral MRE rises from $0.02\%$ to $0.7\%$. Despite this deterioration, the Neural-ODE fits remain substantially more accurate than the analytical Gubser--Rocha benchmark throughout the studied temperature range (Table~\ref{tab:errors}).

This temperature dependence is also reflected in the learned backgrounds. After mapping each candidate to the common $T_H^{\mathrm{ref}}$, the potential $\widehat{qA}_t(v)$ grows from $\sim10^{-4}\,\mathrm{eV}$ at $8\,\mathrm{K}$ to $\sim0.1\,\mathrm{eV}$ at $100\,\mathrm{K}$ and becomes nonmonotonic. The mapped blackening factors depart further from the exact $\mathrm{AdS}_2$ reference, while the curvature of the chosen conformal representative, $\widehat{\mathcal{R}}_2$ in Eq.~(\ref{eq:R2_curvature}), develops irregular profiles and larger near-horizon candidate spread (Figs.~\ref{fig:fha_highT} and~\ref{fig:curvature_lowT}(b)).

Since $qA_t$ shifts the frequency in the Dirac equation, it can help adjust the thermal crossover on one frequency branch while introducing an asymmetry between the two branches. We quantify this asymmetry by the momentum-averaged odd fraction
\begin{equation}
\overline{\mathcal{O}}\equiv
\left\langle
\frac{\sum_{\omega_i>0}\left|\chi_{\mathrm{NODE}}(\omega_i,k_j)-\chi_{\mathrm{NODE}}(-\omega_i,k_j)\right|}
{\sum_{\omega_i>0}\left|\chi_{\mathrm{NODE}}(\omega_i,k_j)+\chi_{\mathrm{NODE}}(-\omega_i,k_j)\right|}
\right\rangle_{k_j},
\label{eq:odd_fraction}
\end{equation}
where the brackets denote the arithmetic average over the sampled momentum grid. For the best $100\,\mathrm{K}$ candidate, the continuation to $\omega>0$ gives $\overline{\mathcal{O}}\approx0.496$, compared with $9.22\times10^{-4}$ at $8\,\mathrm{K}$. This continuation conflicts with the even-frequency structure of the extended-PLL target.

Setting $qA_t\equiv0$ restores frequency symmetry, but the resulting control fit has a larger residual. Fitting both frequency signs also gives a larger residual, whereas relaxing the sign constraint on $qA_t$ improves the negative-frequency fit but leaves pronounced frequency asymmetry.

Taken together, these results indicate that, within the present ansatz and optimization search, we have not found a refined geometry that accurately reproduces the thermal spectral shape while preserving frequency symmetry.

Note that the control fits and numerical convergence checks are documented in Appendix~\ref{app:robustness}, together with the other numerical validation tests and diagnostics.

\begin{table}[htbp]
\caption{The performance of Neural ODEs and the analytical IR Gubser--Rocha model on the normalized extended-PLL targets. In the analytical benchmark, the only tunable parameter is the ratio $q/\mu$ entering $\nu_k=2q\lvert k\rvert/\mu$, fitted separately to each normalized target. The sample labels are adopted from Ref.~\cite{smitMomentumdependentScalingExponents2024}, whose Fig.~1 also provides the PLL parameters $\alpha$ and $\beta$. The nodal $k_F$ values are digitized from Supplementary Fig.~3 of the same reference and rounded to the nearest $0.005\,\text{\AA}^{-1}$. For both the Neural ODE and Gubser--Rocha columns, the loss is defined in Eq.~(\ref{eq:loss}) and the $\chi$ MRE in Eq.~(\ref{eq:mre}).}
\label{tab:errors}\centering
\begin{tabular}{lcccccccc}
\toprule 
\multirow{2}{*}{\textbf{Sample}} & \multirow{2}{*}{\textbf{$T$ (K)}} & \multirow{2}{*}{\textbf{$\alpha$}} & \multirow{2}{*}{\textbf{$\beta$}} & \multirow{2}{*}{\textbf{$k_F$ (\AA$^{-1}$)}} & \multicolumn{2}{c}{\textbf{Neural ODE}} & \multicolumn{2}{c}{\textbf{Gubser--Rocha}} \\
\cmidrule(lr){6-7}\cmidrule(lr){8-9}
& & & & & \textbf{Loss} & \textbf{$\chi$ MRE} & \textbf{Loss} & \textbf{$\chi$ MRE} \\
\midrule 
UD32K & 8 & 0.51 & 3.39 & 0.455 & $2 \times 10^{-8}$ & 0.02\% & $3 \times 10^{-5}$ & 1\% \\
UD32K & 20 & 0.51 & 3.39 & 0.455 & $7 \times 10^{-7}$ & 0.1\% & $9 \times 10^{-4}$ & 5\% \\
UD32K & 60 & 0.51 & 3.39 & 0.455 & $7 \times 10^{-6}$ & 0.4\% & $7 \times 10^{-3}$ & 14\% \\
UD32K & 100 & 0.51 & 3.39 & 0.455 & $3 \times 10^{-5}$ & 0.7\% & $8 \times 10^{-3}$ & 15\% \\
OD23K & 8 & 0.61 & 3.78 & 0.410 & $1 \times 10^{-7}$ & 0.04\% & $8 \times 10^{-5}$ & 2\% \\
OD15K & 8 & 0.74 & 3.66 & 0.410 & $2 \times 10^{-7}$ & 0.06\% & $2 \times 10^{-4}$ & 3\% \\
OD3K & 8 & 0.82 & 4.08 & 0.405 & $3 \times 10^{-7}$ & 0.08\% & $4 \times 10^{-4}$ & 4\% \\
OD0K & 8 & 0.84 & 4.14 & 0.400 & $4 \times 10^{-7}$ & 0.08\% & $5 \times 10^{-4}$ & 4\% \\
\bottomrule
\end{tabular}%
\end{table}

\clearpage
\section{Conclusion and Discussion}\label{sec:conclusion}

We formulated a fermionic inverse problem that learns effective bulk fields entering the probe Dirac dynamics from spectral functions. Methodologically, its radial Neural-ODE inverse architecture builds on the scalar HWRG construction of Ref.~\cite{Ran:2025vat}; physically, the strange-metal target follows the semi-holographic coupling between a boundary electron and a strongly coupled fermionic IR sector~\cite{smitMomentumdependentScalingExponents2024,mauriGaugegravityDualityComes2024}. The extension is formulated directly at the level of the probe Dirac dynamics and therefore does not select a particular bulk action. Its new ingredients are the coupled fermionic radial flow at finite momentum and density, the simultaneous learning of two metric functions and $qA_t$, and a flux-based extraction formula for the chosen massless probe in geometries with conformal-to-$\mathrm{AdS}_2\times\mathbb{R}^2$ asymptotics. The extraction formula can also be used in forward holographic-fermion calculations independently of the Neural-ODE implementation.

After benchmarking the method on the Einstein--Maxwell and Gubser--Rocha models, we applied it to normalized spectra generated analytically by the extended-PLL model for nodal strange-metal phenomenology. For this application, we choose a massless effective probe on the basis of the linear extended-PLL scaling exponent and boundary regularity. This application identifies two analytic redundancies in the fermionic inverse map. First, because the massless probe is insensitive to the conformal factor, the pair $\{f,h\}$ has a coordinate/Weyl redundancy. Fixing the constant-$h$ gauge expresses the result in terms of $\{h_0,F(u),qA_t(u)\}$ but does not determine the conformal factor. Second, even in this gauge, normalized finite-momentum Dirac spectra do not determine the scaled Hawking temperature: regular blackening factors and $qA_t$ profiles related by the temperature map yield identical normalized spectra. A related cutoff--temperature degeneracy arises in the scalar HWRG flow of Ref.~\cite{Ran:2025vat}. The analogy is limited, however: its cutoff is Wilsonian, whereas the radial endpoints here are numerical regulators. After fixing the coordinate/Weyl redundancy and mapping temperature-degenerate candidates to a common-temperature frame, the lowest-loss jointly trained representatives of the low-temperature extended-PLL target have a very small $qA_t$ ($\sim10^{-4}\,\mathrm{eV}$) and blackening factors with only limited profile-level proximity to the exact $\mathrm{AdS}_2$ black-hole form. Their curvatures depart from the $\mathrm{AdS}_2$ value toward the horizon. These profiles characterize the selected fully trainable representatives rather than unique reconstructions of the geometry and gauge potential. Because the conformal factor remains inaccessible to the massless fermionic spectrum, macroscopic thermodynamic quantities such as the entropy density and electronic specific heat require independent input. This distinction is consistent with recent studies of quantum-critical metals: quantum Monte Carlo simulations show that non-Fermi-liquid signatures in the self-energy need not coincide with a broad enhancement of $C/T$~\cite{grossmanSpecificHeatQuantum2021}, while Luttinger--Ward--Eliashberg analyses show that, in the controlled normal-state approximation, the thermal fermionic self-energy cancels from the electronic free-energy contribution, leaving the quantum-critical specific heat encoded in the dressed bosonic propagator rather than directly inferable from single-particle spectra~\cite{zhangFreeEnergySpecificHeat2023}.

Moreover, the normalized construction in the strange-metal application tests the geometric direction identified by Mauri et al.~\cite{mauriGaugegravityDualityComes2024}. At low temperatures, accurate fits with very small $qA_t$ persist across the studied doping range, with only a mild increase in loss toward the overdoped side. At $100\,\mathrm{K}$, the best negative-frequency loss is substantially higher than at $8\,\mathrm{K}$, yet remains well below the analytical Gubser--Rocha benchmark at the same temperature. The learned $qA_t$ is sizable and nonmonotonic, while the mapped blackening factors depart further from the $\mathrm{AdS}_2$ reference and the curvature of the chosen conformal representative becomes irregular, with greater candidate spread near the horizon. In particular, the positive-frequency continuation exhibits pronounced asymmetry, so the candidates found do not yet provide an effective description of the full even-frequency target. 

These findings point to several concrete directions for future work. A natural next step is to relax the assumptions that define the present inverse problem: the strict conformal $\mathrm{AdS}_2\times\mathbb{R}^2$ boundary condition, the chosen minimal massless Dirac probe model, and the clean factorization of the non-holographic contributions $G_0(\omega)$ and $g_k$. Direct fits to measured self-energies would require retaining $G_0$ and $g_k$ and incorporating experimental uncertainties; fits to MDC intensities would additionally require the full electron Green's function together with the experimental response.  Complementary observables, such as macroscopic thermodynamics and Hall-angle scaling, could then be incorporated to constrain the conformal factor and test whether the same effective geometry controls both single-particle spectra and transport. Applying the framework to other strongly correlated systems, including heavy-fermion compounds, and exploring the hidden gravitational dynamics, would further clarify the scope and limitations of holographic and semi-holographic methods in condensed matter physics.

\acknowledgments

We thank Xian-Hui Ge, Cheng-Yuan Lu, Jun Nian, Jian-Pin Wu, and Zhe Yang for helpful discussions. SFW was supported by NSFC grants (Nos.~12675073 and 12311540141). ZYX is supported by the Berlin Quantum Initiative.

\appendix

\section{IR Green's Function and Effective Geometry}\label{app:ir-green}

In this appendix we present a detailed derivation of the IR Green's function for the Gubser--Rocha model at finite temperature, and explicitly identify the corresponding effective bulk geometry.

In the low-temperature, low-energy limits of the Gubser--Rocha model, Ref.~\cite{mauriGaugegravityDualityComes2024} showed that the Dirac equations~(\ref{yz1}) and~(\ref{yz2}) reduce to%
\begin{equation}
\,\partial _{\zeta }%
\begin{pmatrix}
y_{+} \\ 
z_{-}%
\end{pmatrix}%
=-\frac{1}{\zeta \sqrt{1-\zeta ^{2}\delta _{0}^{2}}}%
\begin{pmatrix}
0 & \nu _{k}-\frac{\zeta }{\sqrt{1-\zeta ^{2}\delta _{0}^{2}}} \\ 
\nu _{k}+\frac{\zeta }{\sqrt{1-\zeta ^{2}\delta _{0}^{2}}} & 0%
\end{pmatrix}%
\begin{pmatrix}
y_{+} \\ 
z_{-}%
\end{pmatrix}%
,  \label{yzA1}
\end{equation}%
where%
\begin{equation}
\zeta =2\sqrt{q}\frac{\omega }{3^{3/4}\sqrt{r\mu }},\quad \nu _{k}=\frac{2q\lvert k\rvert%
}{\mu },\quad \delta _{0}=\frac{2\pi T}{\omega },
\end{equation}%
and $\mu$ is the chemical potential. To decouple this system, we introduce the rotated basis $u_{\pm}$,%
\begin{equation}
\begin{pmatrix}
u_{+} \\ 
u_{-}%
\end{pmatrix}%
=\frac{1}{\sqrt{2}}%
\begin{pmatrix}
1 & i \\ 
1 & -i%
\end{pmatrix}%
\begin{pmatrix}
y_{+} \\ 
z_{-}%
\end{pmatrix}%
.
\label{uzf}
\end{equation}%
In this basis the system splits into a second-order equation for $u_{+}$ and an algebraic relation that determines $u_{-}$ in terms of $u_{+}$ and $u_{+}^{\prime}$:%
\begin{align}
\frac{\zeta (i+\zeta )+\nu_{k}^{2}\left[ \left( \delta _{0}\zeta \right) ^{2}-1%
\right] }{\zeta ^{2}\left[ \left( \delta _{0}\zeta \right) ^{2}-1\right] ^{2}%
}u_{+}(\zeta )+\frac{2\left( \delta _{0}\zeta \right) ^{2}-1}{\zeta \left[
\left( \delta _{0}\zeta \right) ^{2}-1\right] }u_{+}^{\prime }(\zeta
)+u_{+}^{\prime \prime }(\zeta ) &=0,  \label{uzODE} \\
\frac{i\nu_{k}}{\zeta \sqrt{1-\left( \delta _{0}\zeta \right) ^{2}}}%
u_{-}(\zeta )+\frac{i}{1-\left( \delta _{0}\zeta \right) ^{2}}u_{+}(\zeta
)+u_{+}^{\prime }(\zeta ) &=0.
\label{ufODE}
\end{align}%
The general solution of the second-order equation~(\ref{uzODE}) can be expressed in terms
of hypergeometric functions.
Imposing the ingoing boundary condition at the black hole horizon
selects the unique retarded solution:
\begin{align}
& u_{+}\sim e^{-2i\pi \nu_{k}}\left( \zeta -\frac{1}{\delta _{0}}\right) ^{%
\frac{i}{2\delta _{0}}}\times \left[ 4^{-\nu_{k}}\zeta ^{-\nu_{k}}\left( \zeta +%
\frac{1}{\delta _{0}}\right) ^{\nu_{k}-\frac{i}{2\delta _{0}}}\,F_{1}\right.  \nonumber \\
& \quad \quad \left. {}-4^{-2\nu_{k}}\left( \frac{\zeta }{\zeta +\frac{1}{%
\delta _{0}}}\right) ^{\nu_{k}}\left( \zeta +\frac{1}{\delta _{0}}\right) ^{-%
\frac{i}{2\delta _{0}}}\frac{\Gamma \left( \frac{1}{2}-\nu_{k}\right) \Gamma
\left( \frac{1}{2}+\nu_{k}-\frac{i}{\delta _{0}}\right) }{\Gamma \left( \frac{1%
}{2}+\nu_{k}\right) \Gamma \left( \frac{1}{2}-\nu_{k}-\frac{i}{\delta _{0}}%
\right) }\,F_{2}\right] ,
\label{uzsolu}
\end{align}%
where%
\begin{align}
F_{1} &=\;_{2}F_{1}\left( -\nu_{k},\frac{1}{2}-\nu_{k}+\frac{i}{\delta _{0}}%
;1-2\nu_{k};\frac{2\zeta }{\zeta +\frac{1}{\delta _{0}}}\right) , \\
F_{2} &=\;_{2}F_{1}\left( \nu_{k},\frac{1}{2}+\nu_{k}+\frac{i}{\delta _{0}}%
;1+2\nu_{k};\frac{2\zeta }{\zeta +\frac{1}{\delta _{0}}}\right) .
\end{align}%
Using Eqs.~(\ref{uzf}), (\ref{ufODE}), and~(\ref{uzsolu}), one can
expand $(y_{+},z_{-})$ near the boundary of the near-horizon region:%
\begin{equation}
\left( 
\begin{array}{c}
y_{+} \\ 
z_{-}%
\end{array}%
\right) =\left( 
\begin{array}{c}
-1 \\ 
1%
\end{array}%
\right) \left( R+J_{\pm }\zeta \right) \left( \delta _{0}\zeta \right)
^{\nu_{k}}+\left( 
\begin{array}{c}
1 \\ 
1%
\end{array}%
\right) \left( S+K_{\pm }\zeta \right) \left( \delta _{0}\zeta \right)
^{-\nu_{k}},
\end{equation}%
where%
\begin{align}
R &=\frac{(1+i)2^{-\frac{1}{2}}2^{-4\nu_{k}}e^{-\frac{1}{2}\pi \left( 4i\nu_{k}+%
\frac{1}{\delta _{0}}\right) }\Gamma \left( \frac{1}{2}-\nu_{k}\right)
\Gamma \left( \frac{1}{2}+\nu_{k}-\frac{i}{\delta _{0}}\right) }{\Gamma \left(
\frac{1}{2}+\nu_{k}\right) \Gamma \left( \frac{1}{2}-\nu_{k}-\frac{i}{\delta _{0}%
}\right) }, \\
S &=(1-i)2^{-\frac{1}{2}}2^{-2\nu_{k}}e^{-\frac{1}{2}\pi \left( 4i\nu_{k}+\frac{1%
}{\delta _{0}}\right) }.
\end{align}%
The IR Green's function $\mathcal{G}_{k}$ is then defined as the ratio%
\begin{equation}
\mathcal{G}_{k}=r^{\nu_{k}}\left( \delta _{0}\zeta \right) ^{2\nu_{k}}\frac{R}{S}%
,
\label{IRGEF}
\end{equation}%
which reproduces Eq.~(47) of Ref.~\cite{mauriGaugegravityDualityComes2024}:%
\begin{equation}
q^{\nu _{k}}\mathcal{G}_{k}/\mu ^{\nu _{k}}=i\left( \frac{q}{3^{3/4}}\frac{%
2\pi T}{\mu }\right) ^{2\nu _{k}}\frac{\Gamma (\frac{1}{2}-\nu _{k})}{\Gamma
(\frac{1}{2}+\nu _{k})}\frac{\Gamma (\frac{1}{2}+\nu _{k}-i\frac{\omega }{%
2\pi T})}{\Gamma (\frac{1}{2}-\nu _{k}-i\frac{\omega }{2\pi T})}.
\label{IRG}
\end{equation}
Note that the imaginary part of Eq.~(\ref{IRG}) remains finite for all positive $\nu_k$, while its real part has poles at positive half-integer values.

We now incorporate these results into our Neural ODE framework. We
first note that Eq.~(\ref{yzA1}) possesses a scaling symmetry under%
\begin{equation}
r\rightarrow sr,\ \omega \rightarrow s\omega ,\ k\rightarrow sk,\
T\rightarrow sT,\ \mu \rightarrow s\mu .
\end{equation}%
This symmetry allows us to rescale all quantities by the horizon radius $r_{h}$ and
directly compare our master equations~(\ref{yz34}) with Eq.~(\ref{yzA1}).
One finds that the IR Green's function~(\ref{IRGEF}) corresponds to the identifications%
\begin{equation}
z=\zeta \delta _{0},\;f(z)=2\pi (T/r_{h}) \left( 1-z^{2}\right) ,\;h(z)=\frac{1}{2\pi
(T/r_{h})}\frac{(\mu/r_{h}) ^{2}}{4q^{2}}.
\end{equation}%
The mass and gauge terms are absent from the leading IR Dirac equation because they have been dropped as subleading contributions. Its realization by the massless effective probe leaves $\Omega(z)$ undetermined by this equation; admissible representatives can be selected using independent regularity conditions. This does not impose a vanishing fermion mass in the original Gubser--Rocha model.
Employing the two scaling symmetries~(\ref{ss1}) and~(\ref{ss2}), the geometry can
be brought to the canonical form%
\begin{equation}
f(z)=1-z^{2} ,\;h(z)=h_{0},  \label{GRfhIR}
\end{equation}%
where $f(z)$ is a special case of Eq. (\ref{eq:F_ads2bh}) with $T_H=1/(2\pi)$ and $h_{0}$ is a constant related to the choice of momentum units.
The resulting geometry is therefore precisely conformal to an $\mathrm{AdS}_{2}$ black hole $\times
\mathbb{R}^{2}$.

\section{Flux-Based Extraction of the Spectral Function}\label{app:flux}

We derive Eq.~(\ref{GXi}) for a massless Dirac field at real frequency and momentum on a real static background. The derivation below uses the momentum branch relevant to our application. We assume regular near-boundary profiles with $f(0)=1$, $h(0)=h_0>0$, and $\nu_k>0$. Our derivation follows the conserved-Dirac-flux approach of Ref.~\cite{GulottaHerzogKaminski2011}. For the conformal-to-$\mathrm{AdS}_2\times\mathbb{R}^2$ asymptotics considered here, we identify the appropriate source projection and normalization, allowing the spectral function to be extracted without isolating the subleading response coefficient.

For $m=0$, the master equations~(\ref{yz34}) become
\begin{equation}
y_{+}^{\prime}=\frac{w-k}{z\sqrt{fh}}\,z_{-},\qquad
z_{-}^{\prime}=-\frac{w+k}{z\sqrt{fh}}\,y_{+},
\label{eq:flux_eom}
\end{equation}
with real coefficients. Define the radial flux
\begin{equation}
\mathcal{J}\equiv\mathrm{Im}\!\left(y_{+}\bar{z}_{-}\right).
\label{eq:flux_def}
\end{equation}
Equation~(\ref{eq:flux_eom}) gives
\begin{equation}
\frac{d}{dz}\!\left(y_{+}\bar{z}_{-}\right)
=y_{+}^{\prime}\bar{z}_{-}+y_{+}\bar{z}_{-}^{\prime}
=\frac{w-k}{z\sqrt{fh}}\,|z_{-}|^{2}-\frac{w+k}{z\sqrt{fh}}\,|y_{+}|^{2},
\end{equation}
whose right-hand side is real. Hence $\mathcal{J}'=0$.

Since the radial system is real, choose complete real source and response solutions $\Phi_S$ and $\Phi_R$ with the boundary normalization
\begin{equation}
\begin{aligned}
\begin{pmatrix}y_+\\z_-\end{pmatrix}
&=S\Phi_S+R\Phi_R,\\
\Phi_S&=z^{-\nu_k}\!\left[\begin{pmatrix}1\\1\end{pmatrix}+o(1)\right],
\qquad
\Phi_R=z^{\nu_k}\!\left[\begin{pmatrix}-1\\1\end{pmatrix}+o(1)\right].
\end{aligned}
\label{eq:flux_full_basis}
\end{equation}
Here $S$ and $R$ are complex coefficients independent of $z$. The $o(1)$ terms include all analytic descendants and any resonant logarithms, and vanish relative to the leading term of their respective branch. Writing the radial system as $\Phi'=M\Phi$, the matrix $M$ has zero trace. Abel's identity and the boundary normalization therefore give
\begin{equation}
\begin{aligned}
W(z)&\equiv\det(\Phi_S,\Phi_R),\qquad
W'(z)=\mathrm{tr}(M)W(z)=0,\\
W(z)&=\lim_{\epsilon\to0}W(\epsilon)
=\det\!\begin{pmatrix}1&-1\\1&1\end{pmatrix}=2.
\end{aligned}
\label{eq:flux_pm}
\end{equation}

To relate $W$ to the flux, write $\Phi_S=(s_1,s_2)^{\mathsf T}$ and $\Phi_R=(r_1,r_2)^{\mathsf T}$, where $s_i$ and $r_i$ are real functions of $z$. Substituting Eq.~(\ref{eq:flux_full_basis}) into Eq.~(\ref{eq:flux_def}) yields
\begin{equation}
\begin{aligned}
\mathcal{J}
&=\mathrm{Im}\!\left[|S|^2s_1s_2+|R|^2r_1r_2+S\bar R\,s_1r_2+R\bar S\,r_1s_2\right]\\
&=(s_1r_2-s_2r_1)\,\mathrm{Im}(S\bar R)
=2\,\mathrm{Im}(S\bar R).
\end{aligned}
\label{eq:flux_coefficients}
\end{equation}
The terms proportional to $|S|^2$ and $|R|^2$ are real; the cross terms use $\mathrm{Im}(R\bar S)=-\mathrm{Im}(S\bar R)$ and $s_1r_2-s_2r_1=W=2$. The in-falling normalization $(y_+,z_-)=(1,i)$ fixes $\mathcal{J}=-1$, so Eq.~(\ref{eq:flux_coefficients}) also guarantees $S\neq0$.

Independently, the leading vectors in Eq.~(\ref{eq:flux_full_basis}) give $y_++z_-=2S z^{-\nu_k}[1+o(1)]$, since the leading response vector has zero component sum. Thus $|y_++z_-|^2=4|S|^2z^{-2\nu_k}[1+o(1)]$. Combining this boundary asymptotic form with the exact flux relation~(\ref{eq:flux_coefficients}), we obtain
\begin{equation}
\mathrm{Im}\,\mathcal{G}_k
=\mathrm{Im}\frac{R}{S}
=-\frac{\mathcal{J}}{2|S|^2}
=-2\mathcal{J}\left.
\frac{z^{-2\nu_{k}}}{\left| y_{+}+z_{-}\right| ^{2}}\right\vert
_{z\rightarrow 0},
\label{eq:flux_extraction_limit}
\end{equation}
which establishes Eq.~(\ref{GXi}). No response coefficient is read from $y_+-z_-$: source descendants can precede the response when $\nu_k>1/2$, and a logarithmic descendant can occur at $\nu_k=1/2$, but neither changes the leading source normalization or the exact flux relation. Equation~(\ref{GXi}) therefore applies for all $\nu_k>0$ on this branch. For the opposite momentum branch, the conserved-flux construction remains valid, with $y_+-z_-$ replacing $y_++z_-$ as the source projection. In the PLL application, the momentum entering the IR Dirac equation after particle--hole conjugation is $2k_F-k>0$ throughout the sampled window.

Finally, $\mathcal{J}=-1$ gives $\mathrm{Im}\,\mathcal{G}_k=1/(2|S|^2)>0$. Together with the positivity of $\mathrm{Im}\,G_{22}$ established in Section~\ref{sec:holo-fermions}, this guarantees $\chi_{\mathrm{NODE}}>0$ in the loss~(\ref{eq:loss}). 

\section{Gauge Fixing and the Blackening Factor}\label{app:F}

For a massless Dirac field, the conformal factor $\Omega(z)$ drops out of the master equations~(\ref{yz34}): the fermionic spectra are sensitive only to the conformal-class metric
\begin{equation}
ds^{2}=\frac{1}{z^{2}}\left[-f(z)\,dt^{2}+\frac{dz^{2}}{f(z)}\right]+h(z)\,(dx^{2}+dy^{2}),
\label{eq:conformal_class}
\end{equation}
together with the charge-weighted gauge potential $qA_t(z)$. The learned pair $\{f,h\}$ is therefore not separately physical. A change of radial coordinate, $z\to u(z)$, accompanied by a Weyl rescaling that preserves the form~(\ref{eq:conformal_class}), can trade content between $f$ and $h$ without changing the fermionic observable. In this appendix we remove this redundancy by a definite gauge choice and identify the gauge-invariant content that survives it.

Since the standard $\mathrm{AdS}_2\times\mathbb{R}^2$ black hole has a constant spatial metric, we use the gauge freedom described above to bring $h(z)$ to its constant boundary value $h_0\equiv h(0)$, placing the learned geometry in the same frame so that the two can be compared directly. Demanding that the line element keep the form~(\ref{eq:conformal_class}) with $h\to h_0$ yields a first-order ODE for the coordinate map $u(z)$, together with the resulting blackening factor $F(u)$,
\begin{equation}
\frac{du}{u^{2}}=\frac{h_0}{h(z)}\,\frac{dz}{z^{2}},\qquad
F(u)=f(z)\,\frac{du}{dz}=\frac{h_0}{h(z)}\left(\frac{u}{z}\right)^{2}f(z).
\label{eq:Fdef}
\end{equation}
We fix the integration constant at the horizon by $u(1)=1$ and integrate Eq.~(\ref{eq:Fdef}) from the horizon to the boundary. Since $f(1)=0$, Eq.~(\ref{eq:Fdef}) gives $F(1)=(h_0/h(1))\,f(1)=0$ automatically as long as $h(1)\neq0$, so the horizon is preserved. Near the boundary, $h(z)\to h_0$ implies $u(z)/z\to1$, hence $u(0)=0$ and $F(0)=1$. The choice $h_0=h(0)$ together with $u(1)=1$ thus fixes the map $u(z)$ and the blackening factor $F(u)$ uniquely.

With $m=0$ the off-diagonal couplings in the master equations~(\ref{yz34}) are built from the two combinations
\begin{equation}
\frac{dz}{z\sqrt{f\,h}}=\frac{du}{u\sqrt{F\,h_0}},\qquad
\frac{dz}{f}=\frac{du}{F},
\label{eq:oneforms}
\end{equation}
which enter multiplied by $k$ and by $\omega+qA_t$, respectively. Direct substitution shows that the map~(\ref{eq:Fdef}) leaves both invariant, while $qA_t$ is carried along as $qA_t(z(u))$. The Dirac system is therefore identical in the $(u,F,h_0)$ and $(z,f,h)$ frames, with the same in-falling condition at the common horizon $z=u=1$, so the solution $(y_+,z_-)$ coincides at corresponding points.

The boundary extraction formula~(\ref{GXi}) is preserved as well. The spinor components and the conserved radial flux coincide at corresponding points, while $h_0$, and hence $\nu_k=\lvert k\rvert/\sqrt{h_0}$, is unchanged. Moreover, $u/z\to1$ near the boundary, so the leading source normalization entering Eq.~(\ref{GXi}) is identical in the two frames. Therefore the flux-based formula derived in Appendix~\ref{app:flux} yields the same spectral function after the transformation. This is precisely the redundancy that renders the separately learned $f(z)$ and $h(z)$ gauge dependent.

After fixing this coordinate/Weyl redundancy, we represent the bulk data in the constant-$h$ frame by $\{h_0,F(u),qA_t(u)\}$. Here $h_0=h(0)$ fixes the IR scaling exponent $\nu_k=\lvert k\rvert/\sqrt{h_0}$, while $F(u)$ gives the blackening profile and in particular the scaled Hawking temperature $T_H=-F'(1)/(4\pi)=-f'(1)/(4\pi)$. The $qA_t$ profile is transported as $qA_t(u)=qA_t(z(u))$ and enters the same Dirac equation. By contrast, thermodynamic quantities depend on the physical spatial metric $g_{xx}=\Omega(z)h(z)$; although this metric is invariant under the coordinate/Weyl rearrangement, the conformal factor $\Omega$ is invisible to the massless fermionic spectra. 

\section{Unit Conventions and Dimensional Analysis}\label{app:units}

In standard holographic calculations, one works in natural units with $L = r_h = 1$. However, comparing bulk quantities with experimental observables requires restoring the appropriate dimensionful constants. In this appendix, we carry out this procedure, establishing a mapping between the dimensionless bulk variables and the dimensionful SI observables.

We begin with the metric ansatz in $r$ coordinates:
\begin{equation}\label{metric r}
ds^{2}=-r^{2}f(r)dt^{2}+\frac{1}{r^{2}f(r)}dr^{2}+h(r)(dx^{2}+dy^{2}),
\end{equation}
where the conformal factor has been stripped off for the chosen massless probe dynamics. Under the coordinate transformation $r \to z=r_h/r$, this becomes
\begin{equation}\label{metric z}
ds^{2}=-\frac{r_h^{2}}{z^{2}}f(z)dt^{2}+\frac{1}{{z^2}f(z)}dz^{2}+h(z)(dx^{2}+dy^{2}).
\end{equation}
Restoring SI units and the AdS radius $L$, Eq.~\eqref{metric z} takes the form

\begin{equation}\label{SI metric z}
    ds^2=-\frac{r_h^2}{z^2L^2}f(z)c^2dt^2+\frac{L^2}{z^2f(z)}dz^2+h(z)(dx^2+dy^2).
\end{equation}
According to the holographic dictionary, the Hawking temperature of the black hole is identified with the field theory temperature. For the metric~\eqref{SI metric z}, the Hawking temperature reads
\begin{equation}
    k_BT=-\frac{\hbar c r_h}{L^2}\frac{f'(1)}{4\pi}.
\end{equation}
Introducing the energy scale $E_t\equiv\hbar c r_h/L^2$, this relation simplifies to
\begin{equation}
    \frac{k_BT}{E_t}=-\frac{f'(1)}{4\pi}.
    \label{TAf}
\end{equation}

We now examine the master equations with SI units restored, following Ref.~\cite{mauriGaugegravityDualityComes2024}. Because $dz/dr=-z^2/r_h<0$, transforming the radial derivative from $r$ to $z$ reverses the overall sign, giving
\begin{align}
\label{yz5}
&\hbar c\sqrt{\frac{g_{xx}}{g_{z z}}}\partial _{z}y_{\pm } =\mp
(\hbar ck-w)z_{\mp },   \\
\label{yz6}
&\hbar c\sqrt{\frac{g_{xx}}{g_{z z}}}\partial _{z}z_{\mp } =\mp
(\hbar ck+w)y_{\pm },\\
\label{u1}
&w=\sqrt{\frac{g_{xx}}{-g_{tt}}}(\hbar c \omega +qcA_{t}).
\end{align}%
Substituting the metric~\eqref{SI metric z} yields

\begin{align}
\label{yz7}
&\frac{\hbar cz}{L}\sqrt{f(z)h(z)}\partial_z y_\pm\pm\left[\hbar c k -\frac{z L}{r_h}\sqrt{\frac{h(z)}{f(z)}}(\hbar\omega+qA_t)\right]z_\mp=0,\\
\label{yz8}
& \frac{\hbar cz}{L}\sqrt{f(z)h(z)}\partial_z z_\mp\pm\left[\hbar c k +\frac{z L}{r_h} \sqrt{\frac{h(z)}{f(z)}}(\hbar\omega+qA_t)\right]y_\pm=0,
\end{align}%
which can be rearranged into the dimensionless form
\begin{align}
    \label{yz9}
    \partial_z y_\pm  \pm\left[\frac{1}{z \sqrt{f(z)h(z)}}\dfrac{\hbar c k}{E_x }-\frac{1}{f(z)}\left(\frac{\hbar\omega}{E_t}+\frac{qA_t}{E_t}\right)\right]z_\mp = 0, \\
    \label{yz10}
    \partial_z z_\mp  \pm\left[\frac{1}{z \sqrt{f(z)h(z)}}\dfrac{\hbar c k}{E_x}+\frac{1}{f(z)}\left(\frac{\hbar\omega}{E_t}+\frac{qA_t}{E_t}\right)\right]y_\pm = 0 ,
\end{align}
where $E_x\equiv\hbar c/L$. Equations~\eqref{yz9} and~\eqref{yz10} reveal that both the fermion energy $\hbar\omega$ and the charge-weighted gauge potential $qA_t$ enter naturally in units of the energy scale $E_t$. According to Eq.~\eqref{TAf}, the combination $\hbar \omega/E_t$ is proportional to the dimensionless ratio $\hbar \omega/(k_B T)$, with the proportionality constant given by the geometric factor $-f'(1)/(4\pi)$. The momentum $\hbar c k$, on the other hand, appears in units of a distinct energy scale $E_x$. However, thanks to the scaling symmetry of the master equations under the simultaneous transformation of $k$ and $h(z)$ in Eq.~\eqref{ss2}, one is free to assign $k$ any convenient experimental unit. Thus the energy variables can be mapped to physical units once $E_t$ is fixed, whereas the momentum conversion retains the arbitrary spatial normalization encoded in $L$; here we adopt \AA$^{-1}$ as a convention.

\section{Geometric Equivalence and Temperature Degeneracy} \label{app:th_degeneracy}

In this appendix, we show that, for a massless Dirac field in the constant-$h$ frame, the normalized fermionic spectral function does not fix the scaled Hawking temperature $T_H$ for any regular blackening factor.\footnote{A related cutoff--temperature degeneracy appears in the zero-momentum $d=0$ scalar HWRG flow underlying Ref.~\cite{Ran:2025vat}. When its finite Wilsonian cutoff is transported, $2\pi\bar T\epsilon/(1-\epsilon)$ is invariant; the SYK cutoff--temperature critical curve follows this invariant, whereas the CFT$_1$ scan at a common small cutoff realizes the mapped equivalence only approximately at finite cutoff. The two settings nevertheless remain distinct: the cutoff is Wilsonian there, while the UV and near-horizon endpoints here are numerical regulators, and the present map acts on finite-momentum Dirac equations rather than a zero-momentum scalar flow.} We keep $qA_t$ in the physical energy units used in Section~\ref{subsec:pll-k}. Appendix~\ref{app:units} then gives the replacement
\begin{equation}
\omega+qA_t\rightarrow T_i\frac{\hbar\omega+qA_{t,i}}{k_BT},
\label{eq:w_temp_map}
\end{equation}
where $T_i=-f_i'(1)/(4\pi)$ is the scaled Hawking temperature of the $i$-th representative, the right-hand side uses the physical frequency and $qA_t$ profile, and $T$ is the boundary temperature in Eq.~(\ref{chidata}).

Let $f_1(u)$ be any regular blackening factor satisfying
\begin{equation}
f_1(0)=1,\qquad f_1(1)=0,\qquad f_1(u)>0\quad(0\leq u<1),\qquad f_1'(1)<0,
\end{equation}
with $h_1=h_0>0$ and $T_1=-f_1'(1)/(4\pi)$. For any $\eta>0$, define
\begin{equation}
    D(v)=1-(1-\eta)v,\qquad
    u=\frac{\eta v}{D(v)},\qquad
    f_2(v)=D(v)^2f_1(u(v)),\qquad h_2=h_0.
    \label{eq:diffeo}
\end{equation}
Since $D(v)>0$ and $du/dv>0$ on $0\leq v\leq1$, this map is monotonic and preserves the boundary and the horizon. Moreover,
\begin{equation}
\begin{aligned}
f_2(0)&=1,\qquad f_2(1)=0,\qquad f_2(v)>0\quad(0\leq v<1),\\
f_2'(1)&=\eta f_1'(1),\qquad T_2=\eta T_1.
\end{aligned}
\end{equation}
Using $du/dv=\eta/D(v)^2$, one obtains
\begin{equation}
    \frac{du}{u\sqrt{f_1(u)}}=\frac{dv}{v\sqrt{f_2(v)}},
    \qquad
    \frac{du}{f_1(u)}=\eta\,\frac{dv}{f_2(v)}.
    \label{eq:temp_oneforms}
\end{equation}
Together with the constant rescaling $t_1=\eta t_2$, the same map gives
\begin{equation}
\frac{1}{u^2}\left[-f_1(u)\,dt_1^2+\frac{du^2}{f_1(u)}\right]
=\frac{1}{v^2}\left[-f_2(v)\,dt_2^2+\frac{dv^2}{f_2(v)}\right],
\end{equation}
so it is a radial diffeomorphism accompanied by a change of dimensionless time unit; constant curvature is not required.

With $m=0$, the master equations contain the two combinations in Eq.~(\ref{eq:temp_oneforms}), multiplied respectively by $k$ and by $\omega+qA_t$. The momentum term is invariant under the map. For the energy term, if $qA_t$ is transported as
\begin{equation}
    qA_{t,2}(v)=qA_{t,1}(u(v)),
    \label{eq:At_transport_temp}
\end{equation}
then Eq.~(\ref{eq:w_temp_map}) gives
\begin{equation}
    T_1\frac{\hbar\omega+qA_{t,1}(u)}{k_BT}\frac{du}{f_1(u)}
    =
    T_2\frac{\hbar\omega+qA_{t,2}(v)}{k_BT}\frac{dv}{f_2(v)}.
\end{equation}
Thus the systems $(T_1,f_1,h_0,qA_{t,1})$ and $(T_2,f_2,h_0,qA_{t,2})$ obey the same master equations at fixed physical $\hbar\omega/(k_BT)$ and $k$, with the same in-falling condition at the mapped horizon.

Near the boundary, $u=\eta v+O(v^2)$. Since a regular $qA_t$ does not affect the exponent $\nu_k=\lvert k\rvert/\sqrt{h_0}$, the leading source combination transforms as
\begin{equation}
    y_++z_-
    =2S u^{-\nu_k}[1+o(1)]
    =2(S\eta^{-\nu_k})v^{-\nu_k}[1+o(1)].
\end{equation}
Thus $\tilde S=S\eta^{-\nu_k}$. Since the conserved radial flux is unchanged by the map, the flux-based formula~(\ref{GXi}) derived in Appendix~\ref{app:flux} gives
\begin{equation}
    \mathrm{Im}\,\tilde{\mathcal{G}}(\omega,k)
    =\eta^{2\nu_k}\mathrm{Im}\,\mathcal{G}(\omega,k).
\end{equation}
The prefactor is independent of $\omega$, and cancels in the normalized ratio:
\begin{equation}
    \tilde{\chi}_{\mathrm{NODE}}(\omega,k)
    =
    \frac{\mathrm{Im}\,\tilde{\mathcal{G}}(\omega,k)}
    {\mathrm{Im}\,\tilde{\mathcal{G}}(\omega_0,k)}
    =
    \frac{\mathrm{Im}\,\mathcal{G}(\omega,k)}
    {\mathrm{Im}\,\mathcal{G}(\omega_0,k)}
    =
    \chi_{\mathrm{NODE}}(\omega,k).
\end{equation}
The temperature degeneracy therefore holds for any regular $f_1$ and any regular $qA_t$ profile transported by Eq.~(\ref{eq:At_transport_temp}), with $qA_t=0$ as the simplest special case. 

\section{Numerical Validation and Diagnostics}\label{app:robustness}

This appendix collects numerical validation tests and diagnostics supporting the extended-PLL results of Section~\ref{subsec:pll-k}. We separate these checks from the main discussion so that the physical conclusions can be read independently of the numerical and restricted-model checks.

\subsection{Input and Sampling Sensitivity}

All tests in this subsection use the UD32K sample at $T=8\,\mathrm{K}$. As a grid-resolution check, we repeated the analysis on a denser $60\times30$ $(\omega,k)$ grid; the results were qualitatively unchanged. To test sensitivity to the digitization of $k_F$, we also repeated the training with $k_F=0.450\,\text{\AA}^{-1}$ instead of $0.455\,\text{\AA}^{-1}$ while keeping the other parameters unchanged. This run gives a best loss of $3\times10^{-8}$ and a $\chi$ MRE of $0.02\%$, and the qualitative features of the learned bulk fields remain unchanged. We also varied the fitted PLL parameters locally by changing $\alpha=0.51\to0.52$ at fixed $\beta=3.39$, and $\beta=3.39\to3.49$ at fixed $\alpha=0.51$. Both fits achieved a best loss of $3\times10^{-8}$ and a $\chi$ MRE of $0.03\%$. Thus the low-temperature fit accuracy is stable under these input perturbations.

\subsection{Low-Temperature Ablations}

The small learned gauge potential at $T=8\,\mathrm{K}$ is not an artifact of the sign-definite form $qA_t=(1-z)n_a(z)^2$. With the sign-unconstrained parametrization $qA_t=(1-z)n_a(z)$, the fit quality is unchanged (loss $\approx4\times10^{-8}$) and $qA_t$ remains small ($<10^{-3}\,\mathrm{eV}$).

To test how much of the improvement over the fixed analytical benchmark can be obtained when only one channel is trainable, we performed two restricted-model fits. Setting $qA_t\equiv0$ while keeping $f$ and $h$ trainable gives a best loss of $2\times10^{-7}$. Conversely, fixing the exact $\mathrm{AdS}_2$ black-hole geometry, $f(z)=1-z$ at $2\pi T_H=1/2$ and $h=(k_F/\alpha)^2$, while allowing only $qA_t$ to vary gives $3\times10^{-7}$. Both losses lie between those of the fixed analytical benchmark and the fully trainable model. Thus either channel alone accounts for most of the improvement, whereas the joint fit provides an additional reduction. These comparisons address channel contributions relative to the fixed benchmark; local profile sensitivity is tested separately in Appendix~\ref{app:frozen_identifiability}.

\subsection{High-Temperature Controls and Numerical Convergence}

At $100\,\mathrm{K}$, fixing $qA_t\equiv0$ raises the best loss to $2\times10^{-3}$ and the $\chi$ MRE to $8\%$. A separate fit of the fully trainable model to both frequency signs gives $1\times10^{-3}$, with a $\chi$ MRE of $6\%$. Repeating the negative-frequency fit on a denser $60\times30$ $(\omega,k)$ grid gives a best loss of $5\times10^{-5}$ and a $\chi$ MRE of $1\%$, leaving the order of magnitude and the qualitative conclusion unchanged. For the lowest-loss retained fit with the sign-unconstrained profile $qA_t=(1-z)n_a(z)$, the negative-frequency loss is $4\times10^{-6}$, with a $\chi$ MRE of $0.3\%$, while continuation to positive frequencies gives $\overline{\mathcal{O}}\approx0.439$. These controls reinforce the tension between fitting the negative-frequency thermal crossover and preserving the even-frequency target.

To test sensitivity to the radial regulators, we used the same tighter cutoffs, $z_{\mathrm{UV}}=10^{-5}$ and $1-z_{\mathrm{IR}}=10^{-7}$, in two complementary checks. First, retraining at these cutoffs yielded best losses of $3\times10^{-8}$ and $8\times10^{-5}$ at $8$ and $100\,\mathrm{K}$, respectively, and therefore did not systematically improve the fits. Second, to isolate forward numerical effects from changes in optimization, we froze the lowest-loss candidates and reevaluated their spectra at the same cutoffs while tightening the ODE tolerances to $\mathrm{rtol}=\mathrm{atol}=10^{-9}$. Relative to the original evaluations, at $8\,\mathrm{K}$ the loss changed from $2\times10^{-8}$ to $8\times10^{-8}$, and the $\chi$ MRE from $0.02\%$ to $0.05\%$; at $100\,\mathrm{K}$, both quantities remained unchanged at the reported precision. Thus neither retraining nor the fixed-weight forward check changes the loss by an order of magnitude or alters the qualitative conclusions.

\subsection{Controlled Synthetic Reconstruction}

As a complementary, fully controlled test, we generated synthetic spectra at $T=8\,\mathrm{K}$ from an exact $\mathrm{AdS}_2\times\mathbb{R}^2$ black hole at the canonical temperature $2\pi T_H^\ast=1$ [$f^\ast(z)=1-z^2$, Eq.~(\ref{GRfhIR})], with constant spatial metric $h^\ast=h_0^\ast=(k_F/\alpha)^2\approx0.796$ (so that $\nu_{k_F}=\alpha=0.51$), carrying the prescribed profile $qA_t^\ast(z)=\mu_0(1-z)$, with boundary value $\mu_0\equiv qA_t(0)\in\{0,0.01,0.1\}\,\mathrm{eV}$. We then inverted them through the same massless, conformal-to-$\mathrm{AdS}_2\times\mathbb{R}^2$ pipeline with the flux-based extraction~(\ref{GXi}) as the extended-PLL analysis, rather than the massive, asymptotically-$\mathrm{AdS}_4$ setting of the Einstein--Maxwell and Gubser--Rocha benchmarks of Section~\ref{sec:examples}. As a result, this also serves as a controlled validation of that pipeline. For the lowest-loss candidates, the boundary value $qA_t(0)$ is recovered accurately in every case: $qA_t(0)\approx 2.88\times10^{-5},\ 0.00998,\ 0.100\,\mathrm{eV}$ for the injected $\mu_0=0,\ 0.01,\ 0.1\,\mathrm{eV}$, all with low loss (Table~\ref{tab:recover_mre}).

Because the normalized spectra do not fix $T_H$, the lowest-loss candidate for each sample yields a freely recovered scaled Hawking temperature:
\begin{equation}
2\pi T_1\approx0.892,\ 0.616,\ 1.11
\quad \text{for}\quad
\mu_0=0,\ 0.01,\ 0.1\,\mathrm{eV}.
\end{equation}
These values generally differ from the injected value $2\pi T_H^\ast=1$. This is the temperature degeneracy of Eqs.~(\ref{eq:diffeo})--(\ref{eq:temp_oneforms}), not a reconstruction error; mapping each reconstruction to the frame with the injected temperature recovers the ground-truth profiles.

\begin{figure}[htbp]
\centering
\includegraphics[width=1.0\textwidth]{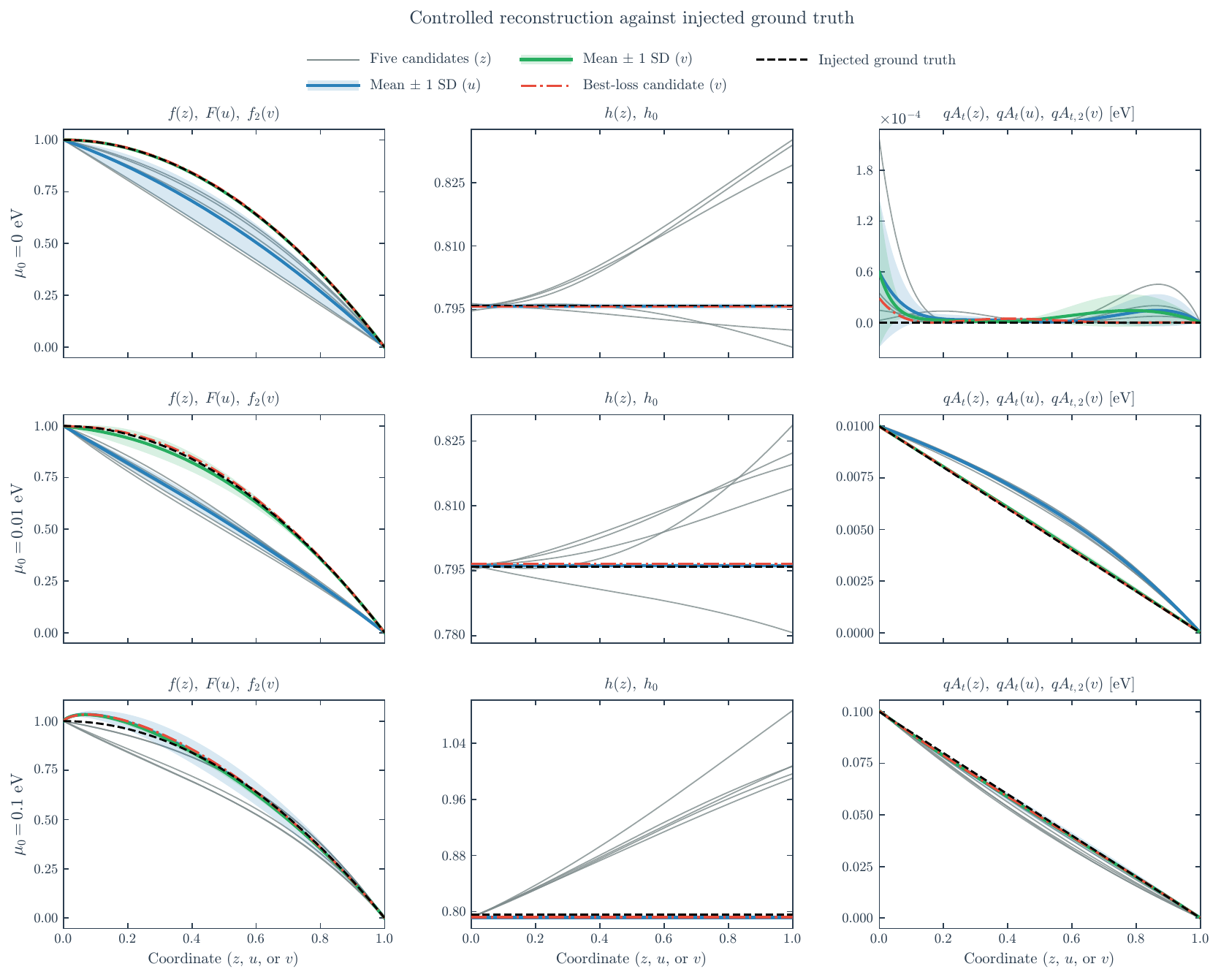}
\caption{Bulk-field reconstruction for the three exact $\mathrm{AdS}_2$ BH samples with $\mu_0=qA_t(0)\in\{0,0.01,0.1\}\,\mathrm{eV}$ (rows); columns are the blackening factor, spatial metric, and $qA_t$ profile. Gray curves show the five raw $z$-frame candidates. Blue and green curves and bands show their means and pointwise 1-SD spreads in the $u$ and $v$ frames, respectively, with both transformations performed separately for each candidate before averaging. Red dash-dotted curves show the lowest-loss candidate in the $v$ frame; black dashed curves show the injected ground truth. For the spatial metric, the $u$- and $v$-frame statistics coincide at the constant $h_0$ and are shown in blue. The $v$-frame profiles are compared directly with the ground truth across all $\mu_0$.}
\label{fig:fha_zuv}
\end{figure}

Figure~\ref{fig:fha_zuv} compares the three bulk functions for all $\mu_0$ in the raw $z$ coordinate, the constant-$h$ coordinate $u$, and the injected-temperature coordinate $v$. Each candidate is mapped separately before the means and 1-SD bands are computed: $z\to u$ fixes the coordinate/Weyl redundancy, while $u\to v$ selects the injected-temperature representative. In the $v$ frame, $f_2(v)$, $h_0$, and $qA_{t,2}(v)$ reproduce their injected profiles, with MREs at the percent level or below for the lowest-loss candidates (Table~\ref{tab:recover_mre}). The reduction of the $qA_t$ MRE from above $10\%$ in the raw $z$ frame to the percent level or below in the $v$ frame confirms that the raw mismatch reflects the temperature degeneracy.

As further tests beyond the exact $\mathrm{AdS}_2$ injections, we applied the same procedure to three additional injections with $f^\ast(z)=(1-z)(1+z^2)$ and $\mathcal{R}_2^\ast(z)=-2+2z^3$ at $(T,\mu_0)=(8\,\mathrm{K},0)$, $(20\,\mathrm{K},0.01\,\mathrm{eV})$, and $(100\,\mathrm{K},0.1\,\mathrm{eV})$. As shown in Table~\ref{tab:recover_mre}, the spectral MREs are below $0.01\%$ and all defined bulk-profile MREs are at or below $2\%$. The reconstruction is therefore not restricted to exact $\mathrm{AdS}_2$ black holes or to the low-temperature regime.

\begin{table}[htbp]
\caption{Controlled reconstruction tests for six synthetic samples using two injected blackening profiles: Profile I is the $\mathrm{AdS}_2$ black hole profile $f_{\mathrm{I}}^\ast(z)=1-z^2$, and Profile II is $f_{\mathrm{II}}^\ast(z)=(1-z)(1+z^2)$. All six use injected $h_0^\ast\approx0.796$ and $qA_t^\ast(z)=\mu_0(1-z)$. The injections are defined in the original $z$ coordinate, whereas the bulk-field MREs are computed after mapping each lowest-loss candidate to the $v$ frame. The loss and spectral MRE are evaluated for the same candidate. The $h_0$ MRE is the scalar relative error $\lvert h_0-h_0^\ast\rvert/\lvert h_0^\ast\rvert$. For the two $\mu_0=0$ cases, the $qA_{t,2}(v)$ MRE is undefined because the reference profile vanishes; the maximum absolute residuals are $3\times10^{-5}\,\mathrm{eV}$ for Profile I and $4\times10^{-5}\,\mathrm{eV}$ for Profile II.}
\label{tab:recover_mre}\centering
\begin{tabular}{lccccc}
\toprule
Profile $(T,\mu_0/\mathrm{eV})$ & Loss & $\chi$ MRE & $f_2(v)$ MRE & $h_0$ MRE & $qA_{t,2}(v)$ MRE \\
\midrule
Profile I $(8\,\mathrm{K},0)$    & $4\times10^{-10}$ & $0.003\%$ & $0.2\%$ & $0.03\%$ & --- \\
Profile I $(8\,\mathrm{K},0.01)$ & $1\times10^{-9}$  & $0.005\%$ & $0.7\%$ & $0.09\%$ & $0.3\%$ \\
Profile I $(8\,\mathrm{K},0.1)$  & $1\times10^{-9}$  & $0.005\%$ & $2\%$   & $0.5\%$  & $1\%$ \\
Profile II $(8\,\mathrm{K},0)$ & $5\times10^{-10}$ & $0.004\%$ & $0.6\%$ & $0.04\%$ & --- \\
Profile II $(20\,\mathrm{K},0.01)$ & $2\times10^{-9}$ & $0.006\%$ & $0.4\%$ & $0.2\%$ & $2\%$ \\
Profile II $(100\,\mathrm{K},0.1)$ & $7\times10^{-10}$ & $0.004\%$ & $0.1\%$ & $0.3\%$ & $0.2\%$ \\
\bottomrule
\end{tabular}
\end{table}

\subsection{Frozen-weight interpolation and local profile sensitivity}\label{app:frozen_identifiability}

We evaluate the lowest-loss UD32K candidate at $8\,\mathrm{K}$ with frozen weights and target parameters. At fixed temperature, the momentum slices of the log PLL target share the same energy dependence and differ only by the factor \(\alpha(k)\). The extra grid points test interpolation of this prescribed analytic shape. At the $39\times19$ cell midpoints, with both energy and momentum absent from the training grid, the spectral MRE is $0.03\%$. The MREs in the $10$--$30$, $30$--$100$, and $100$--$300\,\mathrm{meV}$ intervals are $0.07\%$, $0.02\%$, and $0.02\%$, respectively.

For a local profile test, we fix the constant-$h$ frame and $2\pi T_H^{\mathrm{ref}}=1/2$. We perturb $\log F$ and $qA_t$ using nine bumps proportional to $v(1-v)\exp[-(v-c_j)^2/(2\sigma^2)]$, with $c_j=0.1,\ldots,0.9$ and $\sigma=0.09$, and vary the constant spatial metric value $h_0$. The dimensionless coefficients are normalized so that a unit coefficient produces a peak change of $0.01$ in $\log F$, a peak change of $0.1\,\mathrm{meV}$ in $qA_t$, or a multiplicative change $h_0\to e^{0.01}h_0$, respectively. The perturbations preserve the boundary normalization, horizon position and slope, and $qA_t(1)=0$. We use the singular value decomposition of the Jacobian of $\log\chi$ with respect to the perturbation coefficients to identify its weakest singular direction under this normalization~\cite{hansenDiscreteInverseProblems2010}, then test a finite displacement by solving the full radial Dirac equations. For one admissible displacement, the maximum pointwise relative change in $F$ is $2\%$; the maximum absolute changes in $qA_t$ and $\widehat{\mathcal R}_2$ are $0.157\,\mathrm{meV}$ and $0.735$, respectively. The training loss remains within the retained candidates' loss range, and the midpoint loss increases by less than $4\%$. This local test probes an admissible profile displacement beyond the five retained optimization candidates.

\clearpage
\bibliographystyle{JHEP}
\bibliography{refs}

\end{document}